# AI-coherent data-driven forecasting model for a combined cycle power plant

Mir Sayed Shah Danish[1] (danish@mdanish.me)
Zahra Nazari[2] (znazari@ualberta.ca)
Tomonobu Senjyu[1] (b985542@tec.u-ryukyu.ac.jp)

[1] Department of Electrical and Electronics Engineering, University of the Ryukyus, 1 Senbaru, Nishihara, 903-0213, Okinawa, Japan
[2] Department of Electrical and Computer Engineering, University of Alberta, 116 St & 85 Ave, Edmonton, AB T6G 2R3, Canada

## Abstract

This study investigates the transformation of energy models to align with machine learning requirements as a promising tool for optimizing the operation of combined cycle power plants (CCPPs). By modeling energy production as a function of environmental and control variables, this methodology offers an innovative way to achieve energy-efficient power generation in the context of the data-driven application. This study focuses on developing a thorough AI-coherent modeling approach for CCPP optimization, preferring an interdisciplinary perspective and coming up with a comprehensive, insightful analysis. The proposed numerical model using Broyden Fletcher Goldfarb Shanno (BFGS) algorithm enhances efficiency by simulating various operating scenarios and adjusting optimal parameters, leading to a high yield power generation of 2.23% increase from 452 MW to 462.1 MW by optimizing the environmental factors. This study deals with data-driven modeling based on historical data to make predictions without prior knowledge of the system's parameter, demonstrating several merits in identifying patterns that can be difficult for human analysts to detect, high accuracy when trained on large datasets, and the potential to improve over time with new data. The proposed modeling approach and methodology can be expanded as a valuable tool for forecasting and decision-making in complex energy systems.

**Keywords:** Parameter-based models; data-driven-based models; quasi-Newton method; machine learning method; neural networks; energy efficiency; energy optimization; combined cycle power plant; AI-powered energy system

# 1 Introduction

The growing demand for energy has had a detrimental impact on the environment, both by boosting deforestation and raising carbon and flue gas emissions [1]. Air pollutant refers to any substance that can harm living beings that is mainly caused by the combustion of fossil fuels used in power generation and transportation sectors. $NO_x$ ($NO_x = NO_2 + NO$) is the primary contributor to the atmosphere, leading to environmental problems, e.g., photochemical smog, acid rain, tropospheric ozone, ozone depletion, and global warming, as well as health issues for those exposed to high concentrations of these gases [2]. It is crucial to monitor $NO_x$ and CO pollutants released during combustion operations in power plants [3]. Reports from European countries show that between 2004 and 2020, emissions from large combustion plants in European countries decreased significantly due to policies that set legally binding emission limit values. A decrease in sulfur dioxide ($SO_2$) and dust was observed by 91% and in nitrogen oxides ($NO_x$) by 68% [4]. These emissions are limited by stringent environmental regulations globally; for example, countries of the European Union restrict flue gas emissions, including $NO_x$, CO and dust, which the concentrations of $NO_x$ and CO require to be continuously measured from power plants with a capacity exceeding 100 MW, with $NO_x$ and CO emissions limited to 25 ppmdv when using natural gas as fuel, so flue gas emissions in power plants are monitored by periodic measurements, continuous emission monitoring systems (CEMS) and predictive emission monitoring systems (PEMS) [5].

The shift to climate-friendly sources and the decrease in the use of fossil fuels, particularly coal, contributed to the decline in emissions. Stricter emission limit values and policies to increase the use of renewable or cleaner fuels are expected to drive further decreases in combustion plant emissions in the coming years. In this regard, different machine learning methods, based on various perspectives and algorithms, are utilized in a broad range of applications in energy engineering, e.g., from simple regression and classification problems [6,7] to complex forecasting [8–11] and audio-emotion recognition tasks [12,13]. Distinguishing between indicators, indices, and metrics in an energy model is essential in selecting appropriate tools and techniques for categorization, process identification, evaluation of tangible and non-tangible impacts, and evolutionary confirmation, among others [14]. Sustainability indicators require several criteria, including representativeness and interpretation capability, scientific validity, simplicity, and ease of interpretation, the ability to show trends over time, the ability to give early warning and influence about irreversible trends where possible, sensitivity to change in the environment, society, or economy it is meant to indicate, being based on readily available and adequately documented data, the ability to be updated at regular intervals, and having a target level or guideline against which to compare [15,16].

Meanwhile, energy policy as the main driver of energy sustainability with inappropriate development can lead to inadequate decisions and poor policy prescriptions. Besides, a comprehensive deployment of policy framework and methodologies, integrating AI applications as an inseparable of modern energy infrastructures [17]. Therefore, an in-depth analysis of energy policy development process, procedures, tools, and techniques within a roadmap for the systematic integration of AI platforms in energy sector is known as exigent. In the era of industry 4.0, energy planning and management with a correlative relationship with policies has bestowed nations with socioeconomic prosperity through comprehensive and cross-disciplinary energy demand and supply interactions over time [18]. Climate change mitigation factor has become the pioneering agenda for policymakers in the past two decades. In the next three decades, the importance of AI will be given as an integral part of energy systems automation, monitoring, and control, ensuring reliable supply aligned with techno-economic and sustainability requirements. Several conventions were focused on decreasing global greenhouse gas emissions, with participating parties already having enacted strict environmental regulations, including taxes on carbon emissions [19,20]. Therefore, ensuring access to clean and viable energy and environmentally friendly energy systems requires prioritizing reliability, stability, resiliency, efficiency, exergy and sustainability while considering cost reduction, optimized resources, risk mitigation, and quality maintenance [21].

Tarmanini et al. [22] compared Integrated Moving Average (ARIMA) and Artificial Neural Networks (ANNs) for domestic load forecasting using Mean Absolute Percentage Error (MAPE). They found that ANNs outperformed ARIMA with more accurate results, although they may ignore consumption peaks. Bansal et al. [23] discussed 5 machine learning algorithms by evaluating the algorithms' accuracy, robustness, and reliability, with Long Short-Term Memory (LSTM) and Support Vector Machine (SVM) showing superior performance. The paper also covers the future scope of machine learning algorithms and AI in automation and development, relying on the authors' claims. Mhlanga [24] explored the potential of AI and machine learning to enhance energy production and consumption. The results show that AI and machine learning can optimize consumption, manage the grid, and estimate energy prices and demand. The financial investment in AI and machine learning methods must be robust with performance evaluation measures. Evaluating system measurements and adopting machine learning frameworks is crucial for emerging markets. Xu et al. [25] compared the logistic regression (LR) and the support vector machine (SVM) approaches for the XRF spectral-based classification (SBS) of copper ore samples from Copper Mountain in British Columbia, Canada. PCA and stepwise regression were used for data pre-processing and feature selection, concluding that LR outperformed SVM, with RBF kernel performing better than linear or polynomial kernels, especially on small-sized data. Sarker [26], comprehensively examined deep learning techniques, including a taxonomy for categorizing real-world tasks such as supervised and unsupervised learning. The study found that deep learning can produce high-level data representations from vast raw data, providing solutions to real-world problems. The study emphasized the importance of appropriate data-driven modeling and advanced algorithms trained on the data and knowledge collected to succeed with deep learning. Many studies are focused on data science, artificial intelligence, machine learning, and deep learning methods and tools across various sectors and applications. Some real-world cases are listed in [27–33].

Furthermore, this study aims to provide a comprehensive tutorial on applying machine learning in energy systems focusing on prediction modeling using the Neural Designer to improve the performance of combined cycle power plants (CCPPs). A systematic approach will enable researchers and readers from diverse interdisciplinary domains to better understand the concept of optimization using different machine learning tools and techniques. More importantly, allowing them with it simplified the application process in practice. This study is organized into seven sections. Section 2 deals with the modeling approaches of key characteristics and applications of commonly used tools and techniques retrieved from the extensive literature. Section 3 presents a case study of a real-world combined cycle power plant optimization, its components, indicators used to evaluate performance and efficiency, and the need for an alternative consistent on-demand power supply. Section 4 formulates the system models using the quasi-Newton method optimization algorithm, sets parameters for optimal performance, and introduces a data-driven model. The main topic of optimization is discussed in Section 5, which develops the model architecture and analyzes the optimization throughout the process. The optimization results and the execution of the model are addressed in Section 6, which comprises the formulation, modeling, and flow of the optimization process that results in a numerical model to evaluate optimal operating conditions and efficiencies of the life cycle. Finally, Section 7 concludes the study by highlighting the main contribution of the study.

## 2   Modeling approaches

Integrating AI-based tools, techniques and algorithms, particularly machine learning algorithms, is crucial for contemporary energy policies to meet techno-economic and environmental sustainability constraints. A comprehensive and systematic examination of energy policy development approaches outlines the key characteristics and applications of commonly used approaches, adapted and selected from a comprehensive and competitive reference for sustainability assessment [34]. This information will be valuable for researchers regarding well-understanding energy systems in the context of technical and institutional constraints for proper decision-making on optimization parameters aligned with the objectives [35]. Additionally, understanding a system from the policy and performance components breakdown enables an in-depth and systematic approach to identifying the most influential input variables on the target objectives in the data setting before optimization. Table 1 summarizes the key characteristics and applications of various approaches retrieved, adapted, and selected from a comprehensive reference for sustainability assessment. However, the approaches listed in Table 1 focus on policy development and implementation; meanwhile, these approaches can be deployed for different purposes, including a system analysis in terms of parameters and data performances, and trends.

Table 1 Policy approaches for breaking down an energy system in the context of optimization study input and output variables identification and decision-making [36–39].

| No. | Policy approach | Function and application of the policy approach |
| --- | --- | --- |
| 1 | Quantitative approach | Utilizing quantitative data and statistical techniques, quantifies the costs, benefits, and distributional effects of different energy policy options. |
| 2 | Qualitative approach | Employing nonquantitative data and methods such as interviews and observation comprehend stakeholders' subjective experiences and perceptions. |
| 3 | Interdisciplinary approach | Integrating disciplines such as economics, engineering, and social science, offers a holistic understanding of the impacts of energy policies. |
| 4 | Systems approach | Analyzing energy policies as part of a larger system and investigating the interactions and feedback between different components, realizing the effects of energy policies on other sectors of the economy and society. |
| 5 | Comparative/analysis-orientated approach | Comparing different energy policies in varied contexts to identify similarities, differences, and best practices, identifying the most effective policies for different situations. |
| 6 | Theoretical approach | Utilizing theories from various disciplines such as political science, sociology, and psychology, underlying mechanisms that drive energy policies and their impact. |
| 7 | Empirical approach | Based on data collected from real-world observations or experiments, offering a detailed understanding of how energy policies are implemented and their actual impact. |
| 8 | Participatory approach | Involving various stakeholders in the policymaking process and allowing them to participate in the analysis of energy policies, ensuring that approaches consider the perspectives and needs of different stakeholders. |

| 9 | Risk assessment approach | Evaluating the risk associated with energy policies by taking into account the likelihood and severity of different potential impacts, mitigating potential negative impacts of energy policies. |
|---|---|---|
| 10 | Life-cycle approach | Analyzing energy policies' environmental, economic, and social impacts over their entire life-cycle, recognizing the most sustainable energy policies. |
| 11 | Normative approach | Evaluating energy policies based on pre-established criteria and norms such as ethical principles, human rights, and environmental standards, ensuring energy policies are consistent with broader societal values. |
| 12 | Scenario-based approach | Using different hypothetical scenarios to analyze the potential future effects of energy policies, testing different assumptions, and evaluating energy policies' robustness in different conditions. |
| 13 | Behavioral approach | Examining the psychological, social, and economic factors that influence individuals' and organizations' behavior concerning energy policies, comprehending how people and organizations respond to energy policies, and identifying opportunities for behavior change. |
| 14 | Spatial approach | Analyzing energy policies about geographic space and location, evaluating the spatial distribution of energy policies and their effects on different regions and communities. |
| 15 | Technical approach | Focusing on the technical aspects of energy policies, such as the design and implementation of technology and infrastructure, recommending the most efficient and effective ways to implement energy policies. |
| 16 | Historical approach | Examining energy policies' historical evolution and context, managing how policies have developed over time, and identifying trends and patterns. |
| 17 | International/Cross-national approach | Examining similarities, differences, and best practices across countries and regions in energy policies. |
| 18 | Institutional approach | Approaching legal frameworks and regulations that govern energy policies and how they are enforced. |
| 19 | Dynamic approach | Studying the evolution of energy policies and interactions between different factors over time. |
| 20 | Value-based approach | Evaluating energy policies against societal values such as social justice, environmental protection, and economic development. |

# 3   Case study analysis

A combined cycle power plant (CCPP), particularly those that utilize heat recovery technology, is known for high thermal efficiency, reaching around 66% [40], and low operation and maintenance costs, making them an economically viable option for gas or oil-fired power generation. Furthermore, their integration with environmentally clean gasification systems extends their application to low-cost solid fuel utilization. Their operating flexibility, high reliability and high availability make them suitable for different application types and with a short installation time [41]. Additionally, they present a minimal environmental impact with low stack gas emissions and heat rejection, making them a desirable option for power generation. The CCPP is comprised of several components, including the gas turbine, Heat Recovery Steam Generator (HRSG), steam turbine, generator, cooling tower, air-cooled condenser, control and instrumentation system, natural gas pipeline or fuel storage, electrical grid connection, and intercooler, duct burner, etc. Various objective values are used to evaluate the performance and efficiency (Eq. 1) of a CCPP, such as thermal efficiency, power output, heat rate, availability, reliability, emissions, and so on. Thermal efficiency is one of the most commonly used indicators, calculated as the ratio of net work output to energy input from the fuel source. It represents the amount of useful energy obtained from the fuel source compared to the amount used [42]. The combined cycle process is highly energy-efficient, producing more power with less fuel and minimal environmental impact. The intelligent and reliable grid infrastructure using recent innovations is critical in managing energy from various creation points across distances to end-user and balancing consumers' demand with the available generation resources in real-time [43]. The need for alternative consistent on-demand power is critical as the opportunities for renewables expand.

$$\eta_{th} = \frac{\dot{W}_{net}}{\dot{Q}_{in}} = 1 - \frac{\dot{Q}_{out}}{\dot{Q}_{in}} \quad (1)$$

where, $\dot{W}_{net} = \dot{Q}_{in} - \dot{Q}_{out}$ [W] is system net work rate, $\dot{Q}_{in}$ [W] and $\dot{Q}_{out}$ [W] indicate the input and output heat transfer.

## 3.1 Operation modes

In a CCPP, a synergistic blend of two or more thermodynamic cycles is employed to ensure energy-efficient production and mitigate greenhouse gas emissions [44]. The primary advantages of this system include its controllable output efficiency, high operation reliability, relatively cleaner generation compared to traditional power plants, its ability to swiftly adapt its power output to fluctuating demands for electricity, and a relatively low noise level. However, certain drawbacks are associated with the utilization of CCPPs, the high initial investment requirement, the volatility of natural gas prices as a primary fuel source, a high level of dependence on ambient Temperature and humidity, and the complexity of the system, which may lead to issues such as water contamination and habitat destruction [45]. A combined cycle power plant generates electricity using two cycles. The first cycle produces electricity by using a gas turbine that operates like a jet engine, compressing air mixed with natural gas and igniting it into high-pressure air to spin a turbine attached to a generator. The excess heat from the gas turbine is then used to power a second step of the process, where a specially designed boiler called an HRSG converts water into steam to spin a steam turbine, which is also connected to a generator. The electricity is then fed into the power grid and distributed to the consumers [43]. Below, the detailed working mechanism of a CCPP is presented as follows:

- Compression: A compressor substantially increases density and Temperature by compressing air; at discharge, compressed air with several times density has a temperature that exceeds 300 C.

- Combustion: During the combustion process, the fuel (generally natural gas) is mixed with compressed air and ignited, causing a controlled explosion that produces a high-pressure, high-temperature gas stream. Then passes through the turbine blades, causing them to rotate (mechanical energy) and generate electricity. The exhaust heat from the combustion process is then captured and used to produce steam, which drives a steam turbine to generate additional electricity, thus increasing the overall efficiency of the power generation process in a CCPP.

- Generation: Turbine output utilizes to drive the air compressor and electricity generation, about 60% and 40%, respectively. The turbine inlet is where the compressed air enters, and it is at a lower temperature than the exhaust gas, which is the hot gas that exits the turbine after the combustion process. The turbine exhaust gas is then cooled, and the heat energy is either rejected or recovered depending on the type of power plant. In a single-cycle plant, this thermal energy is typically rejected to the atmosphere without being recovered.

Consequently, a glimpse of the systematic working mechanism of CCPP in Fig. 1 can be sorted as follows [46]:

1. The natural gas pipeline or fuel storage supplies fuel to the gas turbine.
2. The gas turbine generates electricity and also produces exhaust heat.
3. The exhaust heat is then passed through the heat recovery steam generator (HRSG), which is used to heat water and create steam.
4. The steam is then passed through a steam turbine.
5. The steam turbine is connected to a motor/generator which converts the mechanical energy into electrical energy.
6. The cooled exhaust gases from the gas turbine and the steam from the steam turbine are passed through the air-cooled condenser to reject the remaining heat into the atmosphere.
7. The control system monitors and controls the performances and processes within the power plant.

8. The electricity generated by the gas and steam turbines feeds to the power grid using transformer or other facilities if required.

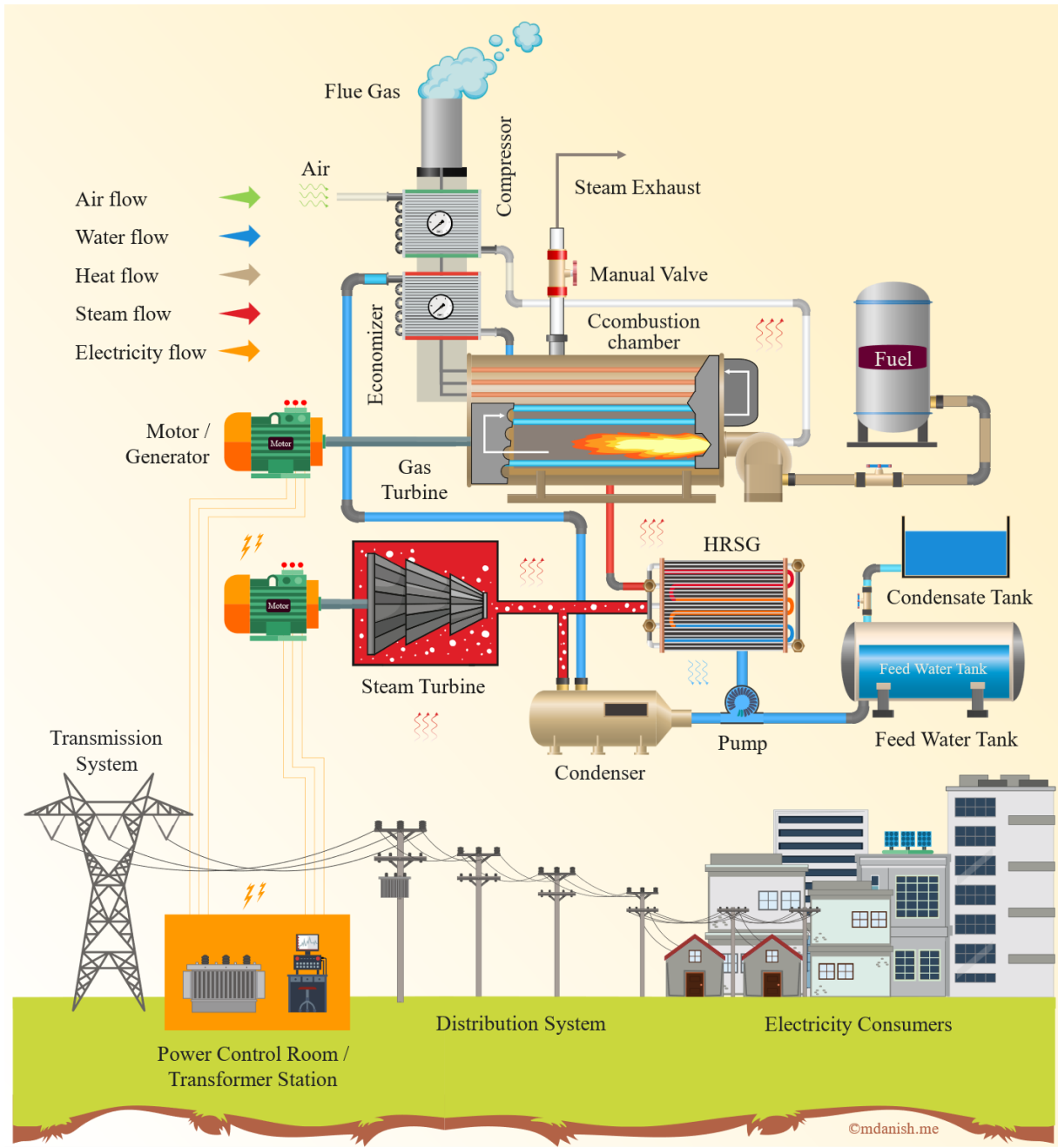

Fig. 1    Components and working mechanism of a typical CCPP.

The combustion turbine of a CCPP operates on the Brayton thermodynamic cycle. The Brayton cycle is a thermodynamic cycle that describes the processes of compression, combustion, expansion, and exhaust in a gas turbine. It can be represented on a temperature-entropy (TS) diagram, a helpful tool for illustrating and analyzing the performance of power plant cycles. On the TS diagram, Temperature (T) is represented on the vertical axis and entropy (s) on the horizontal axis. Entropy is a thermodynamic property of substances that describes the capability of heat energy to work based on temperature. The TS diagram is used to analyze the temperature impacts as it is highlighted with a strong influence variable to the thermodynamic, higher thermal and part-load efficiency of the combustion turbine in the CCPP [47].

The active components determine the operation mode in a combined cycle power plant. When only the gas turbine is in operation, the plant operates in simple cycle mode, while the simultaneous operation of the gas turbine, the heat recovery steam generator (HRSG), and the steam turbine operate in combined cycle mode. In the combined cycle mode, the exhaust heat from the gas turbine is utilized to generate steam, which is subsequently used by the steam turbine to generate additional electricity. This results in an increase in the overall efficiency of the power generation process.

### 3.2 Model and parameters

Several impactful parameters are discussed for an in-depth representation of the proposed study model, the CCPP. Some of these parameters directly impact the output energy in terms of efficiency, while others have an overall systemic impact. The following parameters [48,49] are considered to develop the study model, which closely influences the efficient and reliable operation of a CCPP.

- Temperature (T): Temperature is typically measured as the heat energy of the CCPP at various points throughout the plant, such as the exhaust gas temperature of the gas turbine, the steam temperature in the steam turbine, and the cooling water temperature. Temperature is an important parameter to monitor as it can affect the efficiency and performance of the plant.

- Ambient Pressure (AP): Ambient atmospheric pressure is typically measured at the site of the plant. Ambient pressure can affect the plant's performance, as it can influence the air intake of the gas turbine.

- Relative Humidity (RH): Relative humidity is the amount of moisture in the air compared to the maximum amount of moisture the air can hold at a specific temperature. In a CCPP, relative humidity is typically measured at the project site. High relative humidity can affect the performance of a power station, as it can cause corrosion and increase the cooling water requirements.

- Exhaust Vacuum (V): Exhaust vacuum is the pressure difference between the exhaust gas and the ambient air typically measured at the gas turbine's exhaust. Exhaust vacuum is an important parameter to monitor as it can affect the efficiency and performance of the plant.

- Electrical energy output (EP): Net hourly electrical energy output is the amount of electrical energy generated by a CCPP during a specific hour of operation (MWh).

The principles of mass and energy conservation, as dictated by thermodynamics concepts [50], are used to describe the dynamic behavior of a system. These principles are expressed by Eq. 2, which states that the net transfer rate of energy through heat, work, and mass is equal to the rate of energy change inside a control volume, system, or subsystem, including changes in internal, kinetic, and potential energies. In the case of a steady-flow process, the energy balance for the cycle is given by Eq. 3, which is accompanied by the mass balance equation (Eq. 4).

$$\frac{dE_{sys}}{dt} = \dot{E}_{in} - \dot{E}_{out} \ [J/s] \tag{2}$$

$$\dot{E}_{in} = \dot{E}_{out} \ [W] \tag{3}$$

$$\frac{dm_{sys}}{dt} = \dot{m}_{in} - \dot{m}_{out} \ [kg/s] \tag{4}$$

The basic principles of modeling the compressor and turbine processes by disregarding the changes in kinetic and potential energies, the energy balance equation for a steady flow process can be expressed as follows (Eqs. 5-11) [50]:

$$(q_{in}-q_{out}) + (w_{in}-w_{out}) = h_{ext} - h_{int} \ [J] \tag{5}$$

$$q_{in} = h_B - h_A = c_p(T_B - T_A) \ [J] \tag{6}$$

$$q_{out} = h_C - h_D = c_p(T_C - T_D) \ [J] \tag{7}$$

$$\dot{Q}_{in} = \dot{m}_{FU} \ c_p(T_B - T_A) \ [W] \tag{8}$$

$$\dot{Q}_{out} = \dot{m}_{FU} \ c_p(T_C - T_D) \ [W] \tag{9}$$

$$\frac{T_A}{T_D} = \left(\frac{p_A}{p_D}\right)^{\frac{k-1}{k}} \tag{10}$$

$$\frac{T_B}{T_C} = \left(\frac{p_B}{p_C}\right)^{\frac{k-1}{k}} \tag{11}$$

The turbine and compressor are modeled ideally through equations (4-12) describe the input and output heat transfer ($q_{in}$ and $q_{out}$), works done by the compressor and turbine ($w_{in}$ and $w_{out}$ [J]), pressure ($p$ [Pa]), enthalpy ($h$) [J/mol], temperature ($T$ [K]), gas flow rate ($\dot{m}_{FU}$ [kg/s]), specific heat capacity ($c_p$ [J/kg.K]) specific heat ratio ($k$), and net work rate ($\dot{W}_{net}$ [J]) in a system. The CCPP model shown in Fig. 2 originated based on combined concepts dealing with the element-wise investigation to meet the exact model, consisting of two Gas Turbines, two dual pressure Heat Recovery Steam Generators (HRSG), and a Steam Turbine.

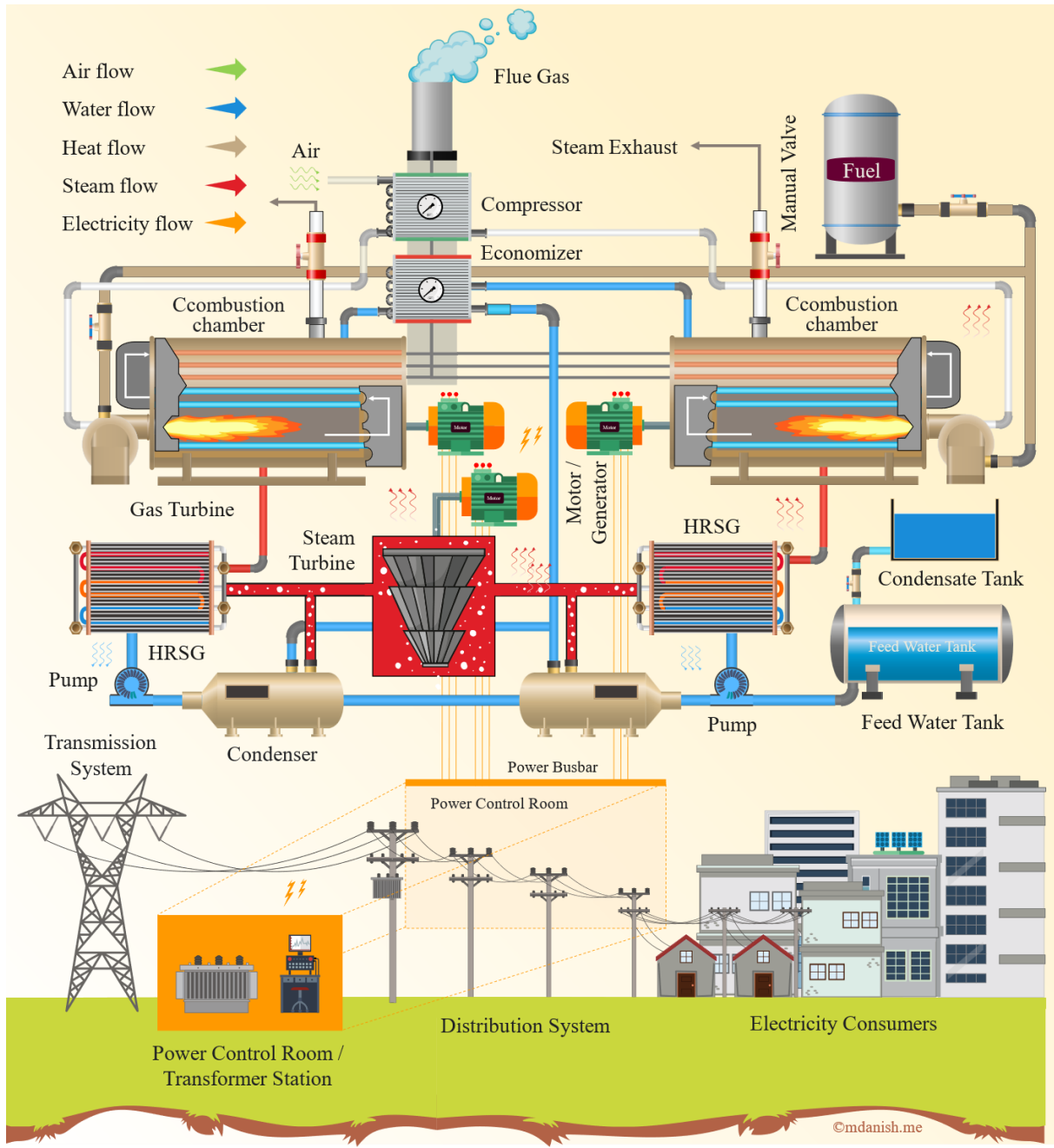

Fig. 2 A CCPP model visualization for this case study consisting 2-Gas Turbines, 1-Steam Turbine, and 2- Heat Recovery Steam Generators (HRSG).

Efficiency analysis and performance evaluation of CCPP requires a comprehensive understanding of various parameters' complex and interrelated behavior. This can be achieved through direct observation, cross-sectional observation, and structured observation approaches, which enable the identification of parameter importance and steady-state evaluation through modeling and real-time monitoring [50–52]. However, meeting the growing demand for smartness in CCPP operations remains a significant challenge [50].

1. Direct observation: from a naturalistic observation perspective, establishing a relationship among these parameters is challenging due to their complex and interrelated behavior. Still, it can be generalized into three relative categories to better understand the parameter (variable) importance in a CCPP model.

- Impact on the system and output efficiency: Temperature, Ambient Pressure, Relative Humidity, Exhaust Vacuum, etc.

- Impact on system performance: Ambient Pressure, Relative Humidity.

- Impact on efficiency and performance: Temperature, Exhaust Vacuum.

2. Cross-sectional observation: In this approach, using historical system performance data, steady-state conditions can be evaluated through mathematical and thermodynamical modeling.
3. Structured observation: In real-time observing the system condition and performances using data-driven models, meeting real-time data transmission, security, scalability, and exchangeability protocols within and beyond the operation cycle. However, the system parameter-based approach is also equipped with online and real-time operation, control, monitoring, etc., facilities, but following the fast-growing trend of smartness remains a considerable challenge.

### 3.3 Data-driven modeling

Data-driven models utilize historical data and machine learning algorithms to identify patterns and make predictions without prior knowledge of the system's physical properties [27]. This is particularly useful for complex, non-linear systems where parameter-based models require accurate data and prior knowledge of the system's behavior [28]. Data-driven models have several advantages, including the ability to identify patterns that may be difficult for human analysts to detect, high accuracy when trained on large datasets, and the potential to improve over time with new data [31]. Overall, data-driven models are valuable for forecasting and decision-making in complex energy systems [21,32,33].

Neural networks (NNs) are machine learning techniques used for approximation, classification, and forecasting. The NNs serve as the approximation or classification model, and the training strategy fits it to the dataset. Model selection algorithms are used to identify the best-performing NNs architecture. The testing analysis compares the NNs outputs with independent targets, and finally, model deployment is used to predict new data. The concept of NNs is modeled after the human brain and consists of three main parts: dendrites, which receive inputs from other neurons; the cell body, where complex functions take place; and the axon, which transmits impulses to other neurons [53]. Neural networks also have an input layer for accepting inputs, hidden layers where complex computations occur (gray-box or black-box), and an output layer responsible for the model's outputs, and the complex computations that occur in the hidden layers are often not fully understood by humans [53].

The data-driven model analysis is broadly categorized into three categories based on the complexity of the applied algorithm and the degree of transparency of the model solution mechanism: white-box, gray-box, and black-box [54]. White-box methods, called "glass-box" or "clear-box" methods, are mathematical models or fully transparent and interpretable algorithms. Humans can easily understand and explain the model's inner workings and decision-making process. These methods use mathematical equations and logic that can be explicitly defined, and the model's architecture is visible to the user. White-box models have the merits of transparency, interpretability, and the ability to make explicit decisions, but the demerits include a lack of robustness and possible overfitting. Gray-box methods, also known as "semi-transparent" or "semi-opaque" methods, are mathematical models or partially transparent algorithms. The decision-making process can be observed, while hidden layers are unobserved. Black-box or "opaque" methods are mathematical models or algorithms that are entirely opaque; humans cannot attend to the model's decision-making process. The most used machine learning methods in energy applications are selected and classified in Table 2 into white-box, gray-box, and black-box techniques.

**Table 2** Classification of the most used machine learning methods in energy applications based on white-box, gray-box, and Black-box methods [55–71].

| No. | White-box methods | Gray-box methods | Black-box methods |
|---|---|---|---|
| 1 | Linear Regression | Artificial Neural Networks | Deep Learning |
| 2 | Decision Trees | Support Vector Machines | Gradient Boosting |
| 3 | Multiple Linear Regression | k-means clustering | XGBoost |
| 4 | Support Vector Regression | Genetic Algorithm | LightGBM |
| 5 | Logistic Regression | ANNs based on Backpropagation | Deep Neural Networks |
| 6 | Naive Bayes | Gradient Boosting | Extreme Learning |

| 7 | K-Nearest Neighbors | Random Forest | |
| 8 | Principal Component Analysis | Deep Belief Networks | |
| 9 | Ridge Regression | Convolutional Artificial Neural Networks | |
| 10 | | Recurrent Artificial Neural Networks | |

Supervised, unsupervised, and reinforcement learning are the subsets of Machin Learning. In supervised learning, the algorithm trains to produce to associate some output with a given input [72]. In supervised learning, the algorithm trains to produce accurate outputs when fed with new datasets, given the knowledge of the associated outputs in online and offline applications [73,74]. Unsupervised learning takes advantage of patterns in the data to create alternatives. Finally, reinforcement learning involves determining the optimal actions to take in response to rewards received from the environment [75].

Supervised learning algorithms aim to accurately associate new input examples with their correct outputs, while unsupervised learning algorithms aim to learn something useful from unlabeled input examples, including tasks such as clustering and dimensionality reduction [76]. Artificial Neural Networks (ANNs) can be categorized into Feedforward Neural Networks (FNNs), Recurrent Neural Networks (RNNs), or multilayer perceptrons (MLPs). FNNs are basic NNs with input, hidden and output layers that data flows in one direction without looping back to the input [77]. They are appropriate for problems with defined input-output relationships. RNNs allow information to circulate in loops and sequential process data like time series or text data. RNNs have hidden units connected to themselves and input from the previous step. They effectively retain information from the previous step and perform well on sequential data tasks such as speech recognition and language translation.

# 4 Model formulation

Analysis of a CCPP is a complex endeavor that requires mindful optimization to accomplish maximum efficiency and system performance. In this study, the quasi-Newton method [78] is applied as an optimization algorithm, unlike Newton's method, which does not need the calculation of the second derivative that offers a more efficient and practical approach for large-scale energy systems, especially from a computation time burdens perspective. In this study, an approximation of the inverse Hessian is computed at each iteration of the algorithm using gradient information. The inverse Hessian approximation method is manipulated to obtain a suitable training rate, while the learning rate method calculates the step for the quasi-Newton training direction. Several parameters are set to ensure optimal performance of the model, including the learning rate tolerance, minimum loss decrease, loss goal, maximum selection error increases, maximum epochs number, maximum time, and so on. Using the quasi-Newton method and intently tuning the parameters, CCPP's optimization results in better performance close to zero correlation, resulting in optimum energy production.

## 4.1 Dataset wrangling

A dataset used for this study is obtained from an actual power plant in a structured format [79], and used to analyze and model the proposed system. A CCPP is a power generation facility that combines gas turbines (GT), steam turbines (ST), and heat recovery steam generators. The electricity generated by the gas and steam turbines is connected in a single cycle. It is transferred from one turbine to another to optimize the overall efficiency of the power plant [80].

### 4.1.1 Data source

The origin of the data in this study encompasses 9,568 data points collected over six years (2006-2011) from a CCPP operating at full-load capacity [79].

### 4.1.2 Variables

The data includes hourly average ambient variables such as Temperature (T) in degrees Celsius, Ambient Pressure (AP) in millibar, Relative Humidity (RH) in percentage, and Exhaust Vacuum (V) in cm Hg, which were used to predict the net hourly electrical energy output (EP) of the power plant in megawatts-hour.

### 4.1.3 Samples (instances)

Fine-tuning an ANNs model necessitates using a dataset comprising a collection of instances, which are judiciously selected, trained, tested, and ignored (if deemed unnecessary), each containing input features and corresponding output labels. The model is trained by adjusting the weights and biases of the network until it can predict the output labels for the desired input features with a high degree of accuracy. The model's performance is subsequently evaluated by comparing its predictions to the actual output labels. Samples play a crucial role in building and validating machine learning models. They are (1) Selected to map the model architecture, (2) Trained to build it, and (3) Tested to validate its functionality.

To ensure that the model is trained and evaluated on diverse instances and to avert overfitting, the dataset is randomly partitioned into three subsets shown in Fig. 3: training, selection, and testing. The training subset, constituting the most substantial portion of the dataset, typically around 60% of the original instances, is used to train the model. The selection subset, comprising a smaller proportion of the dataset, typically around 20% of the original instances, is employed to select the optimal model. The testing subset, comprising around 20% of the original instances, is utilized to evaluate the final model's performance.

#### 4.1.4 Missing (unnecessary values)

The dataset has complete information for the variables in question, and no missing values exist.

#### 4.1.5 Data set objectives

It used for the specific tasks or goals for which the dataset is being used, such as building a predictive model or analyzing trends.

### 4.2 Dataset curation

In a dataset, columns of attributes (features) can be categorized as (1) input, (2) target, or (3) unused data based on their numeric type and sorted using various scaling methods (MinMaxScaler, StandardScaler, RobustScaler, Normalizer, QuantileTransformer with normal distribution). In this study, 4 columns are used as inputs and 1 as the target attribute (Fig. 3).

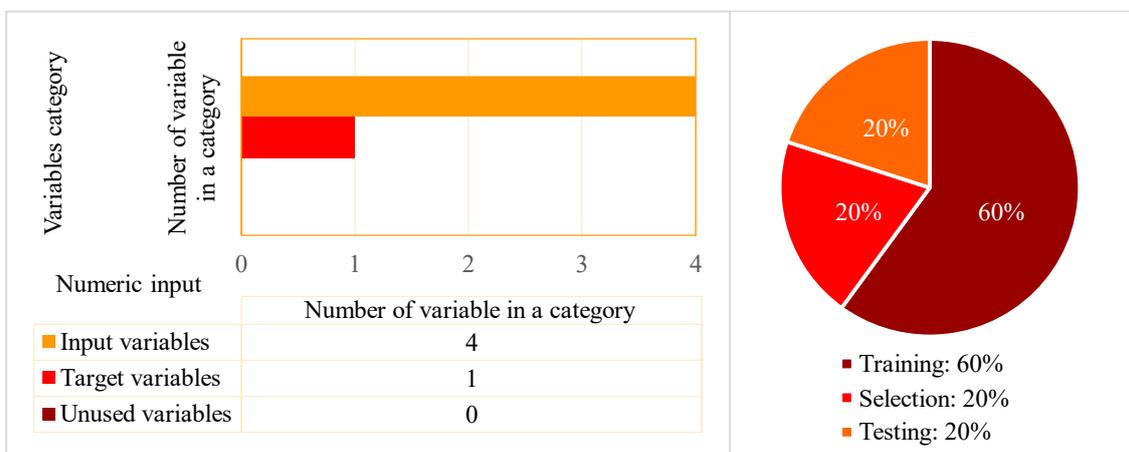

Fig. 3   The system dataset's distribution of 9,568 samples with 4 input variables, 1 target variable, and 0 unused variables through allocation percentages of training, selection, and testing, with no missing values.

Developing an accurate machine learning model requires a thorough understanding of the statistical properties of the data and identifying any anomalies or inaccuracies that may hinder the model's performance. Figure 4 presents the data statistics, including minimum and maximum values, averages, and standard deviations, that ensure the data is clean and free from errors or outliers that may negatively impact the model's accuracy. Statistical variables are used to check for the correctness of every single variable that represents in terms of (1) Minimum (the highest value of a variable), (2) Maximum (the lowest value of a variable), (3) Mean (the average value of a variable), and (4) Deviation (the spread of the variable's values) in the dataset. These measures help understand the distribution and range of the data and identify outliers for designing a model.

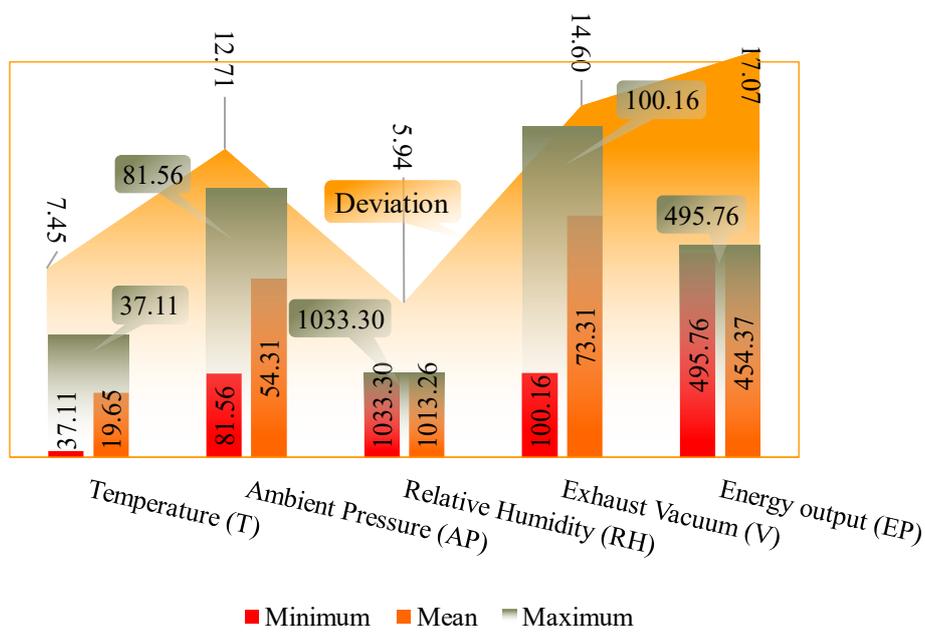

Fig. 4   The statistical information of minimum, mean, and maximum parameter-based values of the model performance.

## 4.3  Data distribution

Selecting a scaler for further dataset processing is essential in an NNs method. It is a pre-processing technique that normalizes the input data before it is fed into the network. Normalization aims to scale the input data to a typical range of values. There is no specific guideline for selecting types of scalers; it depends on the type and distribution of the data. Choosing the appropriate scaler for the dataset to achieve optimum network performance is crucial. Normalizing input data improves ANNs' performance and accelerates convergence. Min-max scaler scales data between a given range, usually 0-1, for known ranges. Standard scaler standardizes data for normal distributions by subtracting the mean and dividing by the standard deviation. Normalization is applied to training, validation, and testing data for consistent comparison. Some scalers used in ANNs are briefly explained as follows [81–83]:

1- No scaling: In this approach, data is utilized in raw form and refrains from scaling to the input data.

2- Minimum/Maximum: This scaler normalizes the input data by scaling it within a defined range, usually between 0 and 1.

3- Mean Standard Division (MSD): This approach applies for standardizing the data by subtracting the mean and dividing the result by the standard deviation; this scaler is beneficial for datasets with a normal distribution, such as real-valued data. This technique is also known as Z-score standardization.

4- Logarithm: This scaler is employed to handle data with a skewed distribution or to bring down the scale for an extensive range of datasets; it applies a logarithmic transformation to the input data. The logarithm can be base ten or a natural logarithm (base e). This approach is beneficial for datasets that contain large values.

Upon examination of the data distribution, as depicted in Fig. 5, the Mean Standard Division scaler is preferred to handle outliers. It brings all features to the same scale, which is essential for optimization algorithms to converge appropriately, ensuring that the data is well-conditioned for training and improving the NNs' performance. Mean Standard Division (MSD) scaling is a normalization method used in machine learning to standardize input variables into a Gaussian distribution with mean 0 and standard deviation 1. The formula for MSD scaling is given in Eqs. 12-14; $\mu$ is calculated as the average of all values, and $\sigma$ measures the spread of data around the mean and is calculated as the square root of the variance. This normalization ensures all input data has an equal impact on the model.

$$x' = (x - \mu) / \sigma \tag{12}$$

$$\mu = x_1 + x_2 + x_3 + \cdots + x_n/n \tag{13}$$

$$\sigma = \sqrt{((x_1 - \mu)^2 + (x_2 - \mu)^2 + (x_n - \mu)^2)/n} \tag{14}$$

Where, $x$ is the original data, $x'$ is the scaled data, $\mu$ is the mean of the data, and $\sigma$ is the standard deviation of the data.

Column distribution is a summary of values in a specific dataset column and provides information on the spread and center of the data. Bin numbers refer to the intervals into which data is divided and can be visualized as a histogram. Indices are numerical labels that identify specific elements in a data set, e.g., binned data points in histograms. The distribution of various variables in Fig. 5 shows the frequency distribution of the variables across their range, with the x-axis representing the bin numbers and the y-axis representing the frequencies. A uniform distribution of all variables is preferred for better model quality, while an irregular distribution results in poor model quality. Temperature distribution has a maximum frequency of 12.14% at 24.2 and a minimum of 0.10% at 35.9, with a range of 1.81 to 37.1, first quartile of 13.5, median of 20.3, and third quartile of 25.7. The Exhaust Vacuum distribution has a maximum frequency of 21.31% at 42.2 and a minimum of 0% at 31, with a range of 25.4 to 81.6, first quartile of 41.7, median of 52.1, and third quartile of 66.5. The Ambient Pressure distribution has a maximum frequency of 18.27% at $1.01 \times 10^3$ and a minimum of 0.10% at 994, with a range of 993 to $1.03 \times 10^3$, first quartile of $1.01 \times 10^3$, median of $1.01 \times 10^3$, and third quartile of $1.02 \times 10^3$. The Relative Humidity distribution has a maximum frequency of 13.03% at 82.8 and a minimum of 0.09% at 28, with a range of 25.6 to 100, first quartile of 63.3, median of 75, and third quartile of 84.8. The Energy Output distribution has a maximum frequency of 12.52% at 438 and a minimum of 0.18% at 423, with a range of 420 to 496, first quartile of 440, median of 452, and third quartile of 468.

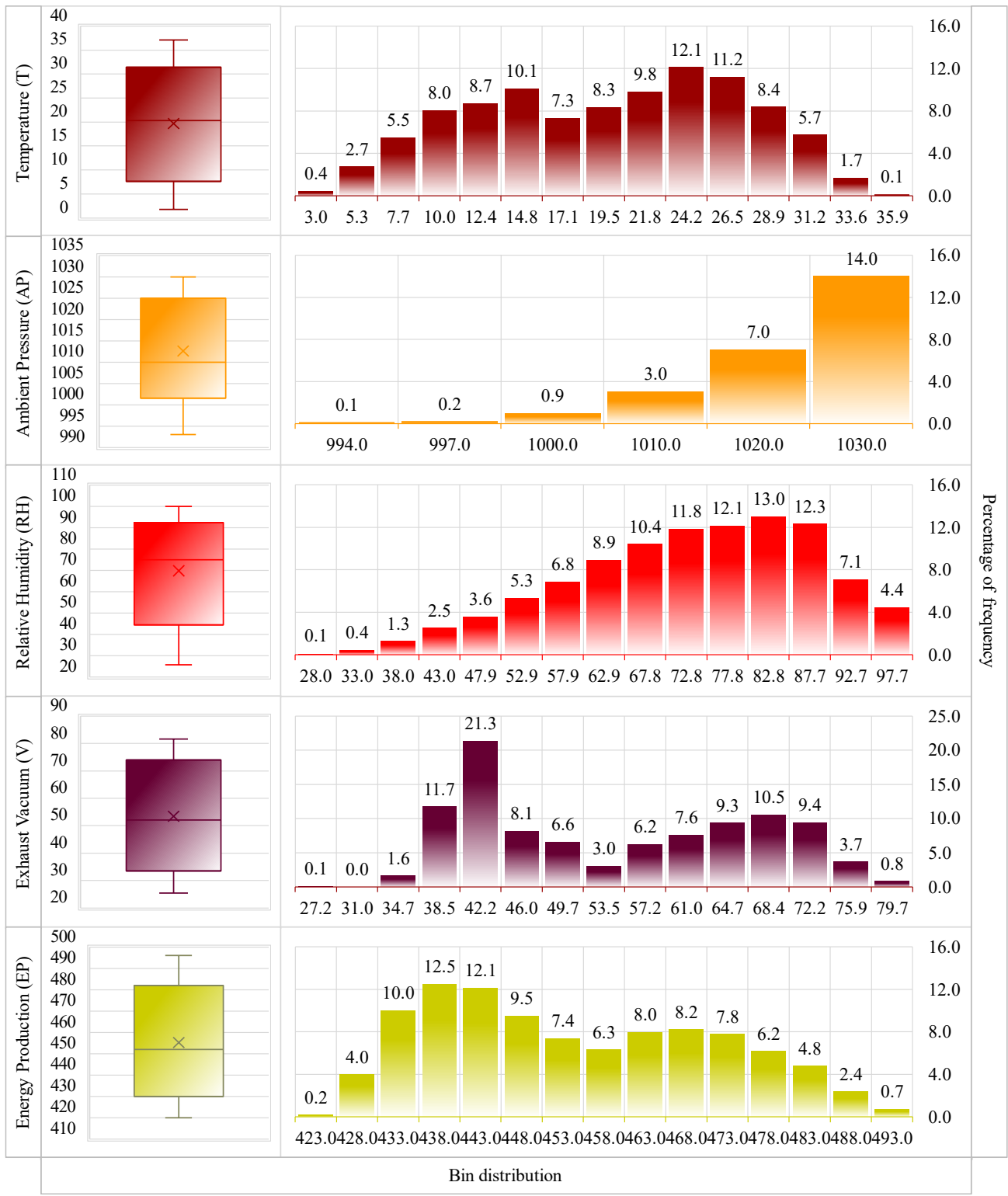

Fig. 5 Visualization of the column spread and bin centers numbers for the frequency distribution of system analysis parameters (instances), with their respective maximum and minimum frequencies, ranges, quartiles, and medians, highlighting the preference for a uniform distribution of optimum model quality.

The Energy Output vs. Inputs graphs provides valuable insights into the relationship between the target and input variables in the data set. In Fig. 6, the regression line is used to measure the strength and direction of the relationship between the two variables. The plotted regression lines of different types, such as exponential or power, can provide information on the nature of the relationship between the input and target variables. The correlation values ranging from -0.948255 to 0.520263 for Temperature, Exhuase Vacuum, Ambient Pressure, and Relative Humidity show the degree of correlation between the input and target variables. A correlation value of 1 indicates a perfect positive correlation, while a value of -1 represents a perfect negative correlation. In this case, the correlation values range from strongly negative to moderate positive correlations, indicating that some input variables significantly impact the target variable.

By visualizing 1000 random samples of the entire set of used samples, the chart provides a better understanding of the interdependencies between the input and target variables. This allows us to identify any data patterns or trends that may help improve the system's performance. The Energy Output vs. Inputs graphs with the depicted regression line and 1000 random samples visualization comprehensively analyze the interdependency of the system's input variables on the output generated. By understanding these interdependencies, we can identify the factors that affect the system's performance and optimize it accordingly.

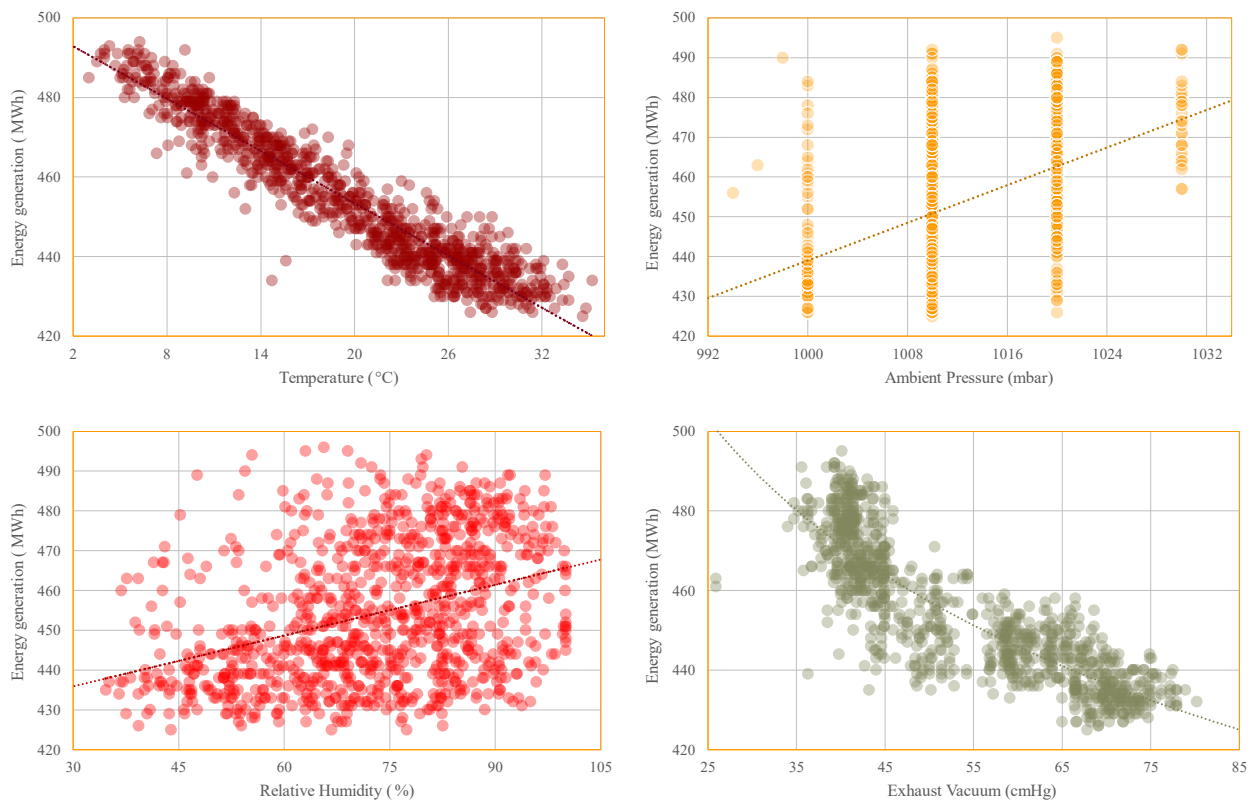

Fig. 6    Influence of interdependency of the system input variables on the output generated in the presence of regressions and 1000 random sampling visualization.

A negative correlation value (-0.948255 for Temperature) indicates that the Energy Output decreases as the input variable (Temperature) increases. On the other hand, a positive correlation value (0.520263 for Ambient Pressure) indicates that as the input variable (Ambient Pressure) increases, the Energy Output also increases. The closer the correlation value is to 1 or -1, the stronger the relationship between the Energy Output and the input variable. For example, the correlation value of -0.948255 for the regression line of type exponential between the target and input temperature variables suggests a strong negative relationship between the two. On the other hand, the correlation value of 0.520263 for the regression line of type exponential between the target and input relative humidity variables suggests a weak positive relationship. The Energy Output vs. Inputs charts provides valuable insights into the relationships between the target and input variables in the data set. This information identifies dependencies between the targets and inputs, which can inform decision-making processes related to energy production and consumption.

## 4.4 Data correlations

Correlation measures the linear relationship between two variables. It ranges from -1 to 1, where -1 indicates a perfect negative correlation, 1 indicates a perfect positive correlation, and 0 shows no correlation. Maximal input correlations refer to the maximum correlation between input and target variables. The Spearman inputs correlations refer to the maximum value of Spearman's correlation between input variables and the target variable. Input correlations refer to the correlation between input variables and the target variable, while Spearman inputs correlations refer to Spearman's correlation between input variables and the target variable.

The given Maximal correlation data among the top 4 inputs in Fig. 7 describes a linear relationship between the input variables and the energy output of a system. The correlation of the inputs in Fig. 7 (a) shows that Temperature has a strong positive correlation with Exhaust Vacuum (85%) and negative correlations with Relative Humidity (54%) and Ambient Pressure (50%). Exhaust Vacuum has a negative correlation with Ambient Pressure (41%).

The Maximal correlation among all input variables in Fig. 7 (b and d) shows the correlation values between each pair of input variables, where Temperature has a strong positive correlation with Exhaust Vacuum (85%) and negative correlations with Ambient Pressure (50%) and Relative Humidity (54%). Exhaust Vacuum has a positive correlation with Temperature (85%) and negative correlations with Ambient Pressure (41%) and Relative Humidity (31%). Ambient Pressure has a weak positive correlation with Relative Humidity (10%). The variation of the spearman correlation in comparing variable sets is given in Fig. 8.

The Spearman of the top 4 greatest inputs correlations in Fig. 7 (c) are similar to the correlations of the Maximal input, where Temperature has a strong positive correlation with Exhaust Vacuum (85%) and negative correlations with Relative Humidity (54%) and Ambient Pressure (52%). Exhaust Vacuum has a negative correlation with Ambient Pressure (43%).

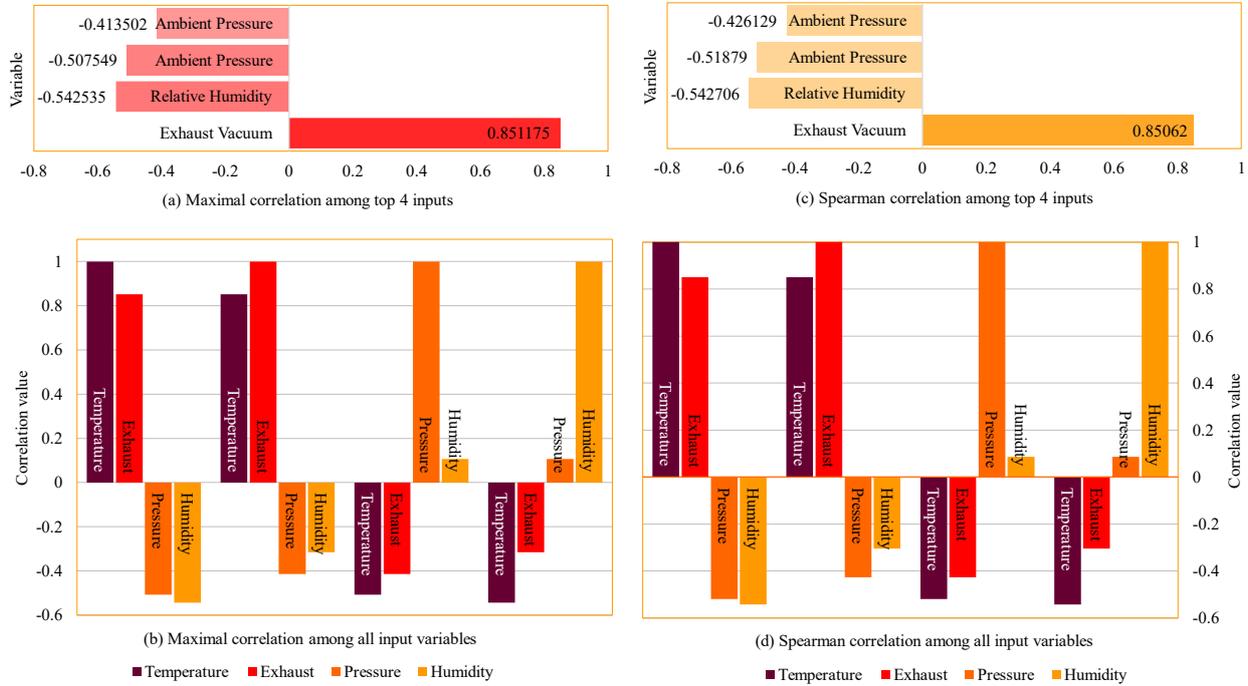

Fig. 7   Interdependency analysis of the input variables and target objective is depicted as (a) the Maximal correlation among the top 4 input variables, (b) the Maximal correlations among all inputs, (c) The Spearman correlations of the top 4 inputs, and (d) the Spearman correlation among all input variables.

Figure 8 exhibits the percentage difference between the Maximal and Spearman correlations for the four input variables. The diagonal values are all zero because the Maximal and Spearman correlations are identical when comparing a variable with itself. The most considerable percentage difference is 1.92%, between Ambient Pressure and Relative Humidity.

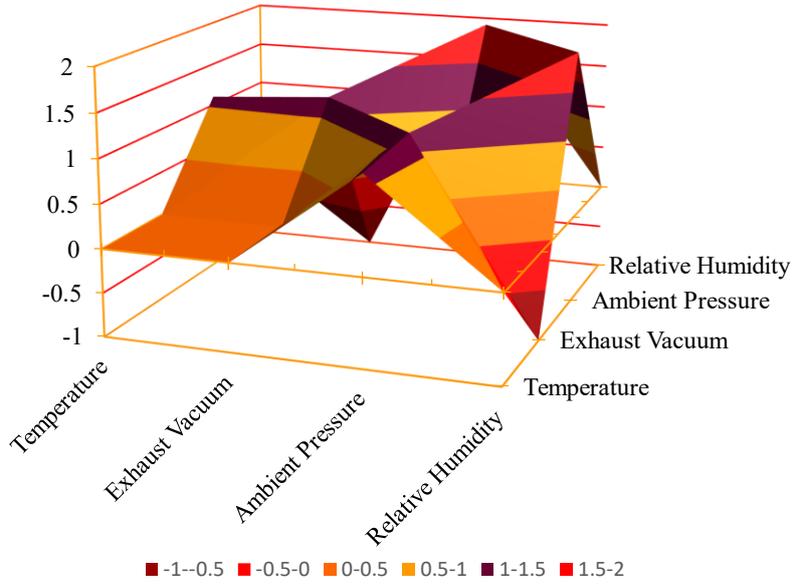

Fig. 8    Percentage variations of Maximal and Spearman correlation coefficients calculation methods of each pair of input variables.

The energy output Pearson correlations in Fig. 9 (a) illustrate dependency on the input columns with the greatest correlation of the dataset, in which Temperature and Exhaust Vacuum have strong negative correlations with energy output (95% and 88%, respectively). Ambient Pressure has a moderate positive correlation with energy output (52%), and Relative Humidity has a weak positive correlation (39%). Pearson's correlation coefficient (r) is given as follows (Eq. 15) [84]:

$$r = \frac{\sum_{i=1}^{n}(\hat{x}_i - \bar{\hat{x}}_i)(x_i - \bar{x}_i)}{\sqrt{\sum_{i=1}^{n}(x_i - \bar{\hat{x}}_i)^2 \sum_{i=1}^{n}(\hat{x}_i - x_i)^2}} \qquad (15)$$

where $x_i$ and $\hat{x}_i$ are individual data points, $\bar{\hat{x}}_i$ and $\hat{x}_i$ are the means of $x_i$ and $\bar{x}_i$, respectively, and n is the number of data points.

The Spearman inputs correlations, as shown in Fig. 9 (b) are similar to the above Pearson in Fig. 9 (a), where Temperature has a strong positive correlation with Exhaust Vacuum (85%) and negative correlations with Ambient Pressure (52%) and Relative Humidity (54%). Exhaust Vacuum has a positive correlation with Temperature (85%) and negative correlations with Ambient Pressure (43%) and Relative Humidity (31%). Ambient Pressure has a weak positive correlation with Relative Humidity (9%).

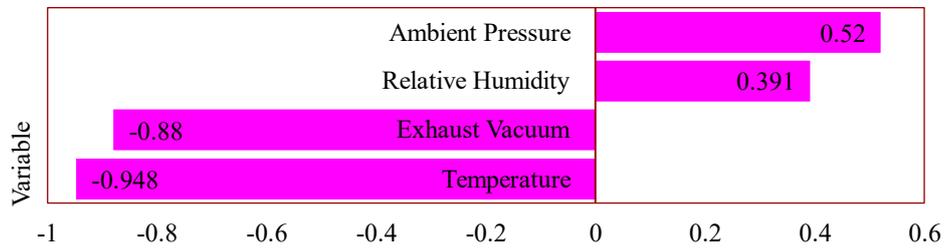

(a) Target value dependency to the individual input variables - Pearson method

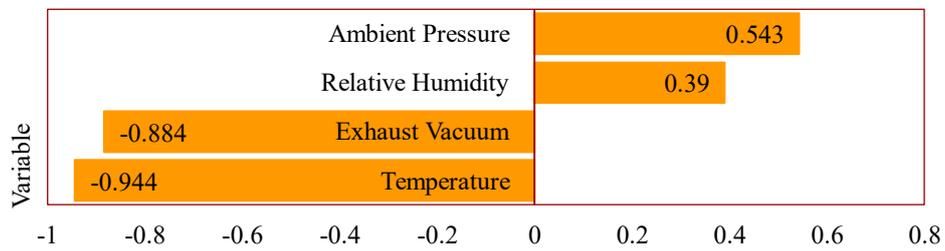

(b) Target value dependency to the individual input variables - Spearman method

Fig. 9  The input variables and target objective correlation coefficient analysis consider single target value dependency on the single input in the dataset using Pearson and Spearman correlation methods.

Spearman's correlation is a non-parametric method that measures the monotonic relationship between two variables and is not limited to linear relationships. The Spearman correlation coefficient (ρ) is given by the following equation (Eq. 16) [85]:

$$\rho = 1 - \frac{6 \sum d_i^2}{n(n^2 - 1)} \qquad (16)$$

where $\rho$ = Spearman's rank correlation, $d_i$ is the difference between the ranks of observation, and $n$ is the number of data points.

The finding in Fig. 10 shows that Ambient Pressure and Relative Humidity have moderate and weak positive correlations with energy production, respectively, while Exhaust Vacuum and Temperature have solid and robust negative correlations with production, respectively. The Spearman correlation coefficients indicate slightly stronger monotonic relationships between Ambient Pressure and energy production and slightly weaker relationships between Relative

Humidity and energy generation compared to the Pearson correlation coefficients. The results suggest that Exhaust Vacuum and Temperature are the most important input variables for predicting the objective target. At the same time, an increase in Ambient Pressure and Relative Humidity may lead to an increase in generation. Still, the relationship is weaker than between Exhaust Vacuum and Temperature and target value.

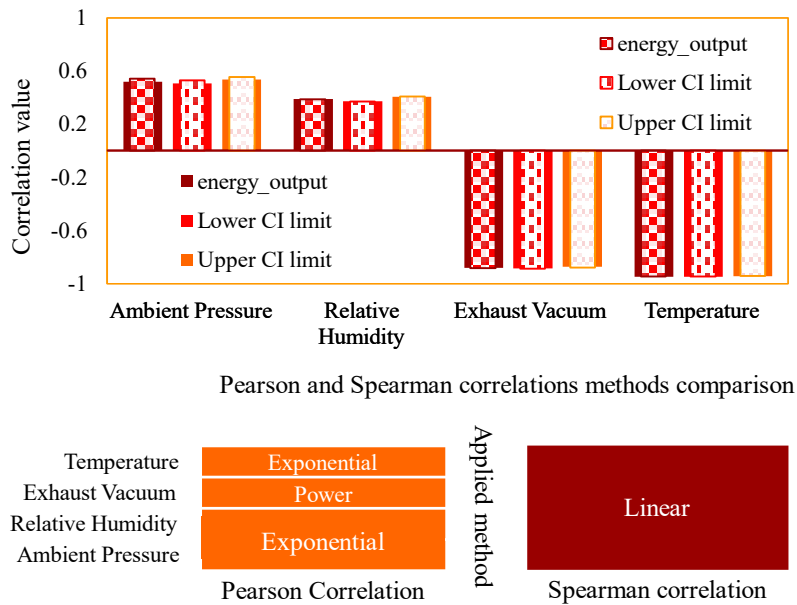

Fig. 10   A comparative evaluation of the correlation coefficients and the degree of linear (Pearson) and monotonic (Spearman) relationship between individual input variables and the target values.

The report indicates a close conformity of the Pearson and Spearman correlation coefficients and their corresponding 95% confidence intervals for the relationship between input and target variables. In this scenario, all variables exhibit some form of correlation with no uncorrelated variables in the data set. These analyses provide insight into the strength and direction of the relationships between input variables and output system variables to define which variables impact system performances most and inform the development of predictive models.

# 5   Model optimization

The optimization technique in this study consists of various building blocks [72], including an input layer to receive the input data, hidden layers that use non-linear transformations to process the data, an output layer to generate the final prediction, activation functions to introduce non-linearity into the network, weights to control the strength of connections between nodes, biases to shift activation functions, a loss function to measure the difference between predicted and actual outputs, an optimizer to update weights and biases based on the gradient of the loss function, and forward and backward passes to compute the output and gradients of the loss function, respectively, used to update the parameters during training. Optimization deployment in this study follows a specific order of building blocks: the input layer, forward pass, loss function, backward pass, optimizer, and output layer. The input layer receives input data, followed by the forward pass, which computes the network's output using activation functions and matrix operations. The loss function measures the difference between predicted and actual outputs, and the backward pass computes gradients of the loss function using backpropagation. The optimizer updates the weights and biases based on the loss function's gradient, and the output layer generates the final prediction of the plant power generation.

## 5.1 Model architecture

Four inputs are fed to the model, normalizing the inputs with the size of the scaling layer 4 (inputs) used to scale the inputs with the minimum, maximum, mean, and standard deviation. The number of perceptron layers is two, with 4 and 3 inputs supported by 3 and 1 neuron numbers, respectively, which are fundamental units in the neural network responsible for making predictions of the size of each layer and its corresponding activation function. The size of the upscaling layer is 1 (output) expands the output from the previous layer to the final results of the model with the values used for scaling the inputs. Scaling and upscaling layers concerning the input variables and energy production statistics are figured out in Figs. 11-12. The Hyperbolic Tangent ($\sigma(\gamma)$) activation function (like logistic sigmoid) with the advantage of mapping positive - strongly positive, negative inputs - strongly negative and the zero inputs - near zero in the activation graph based on the feedforward multilayer perceptron (MLP) is applied, which is given in the follows equation (Eq. 17) [86,87].

$$\sigma(\gamma) = \frac{e^\gamma - e^{-\gamma}}{e^\gamma + e^{-\gamma°}} \tag{17}$$

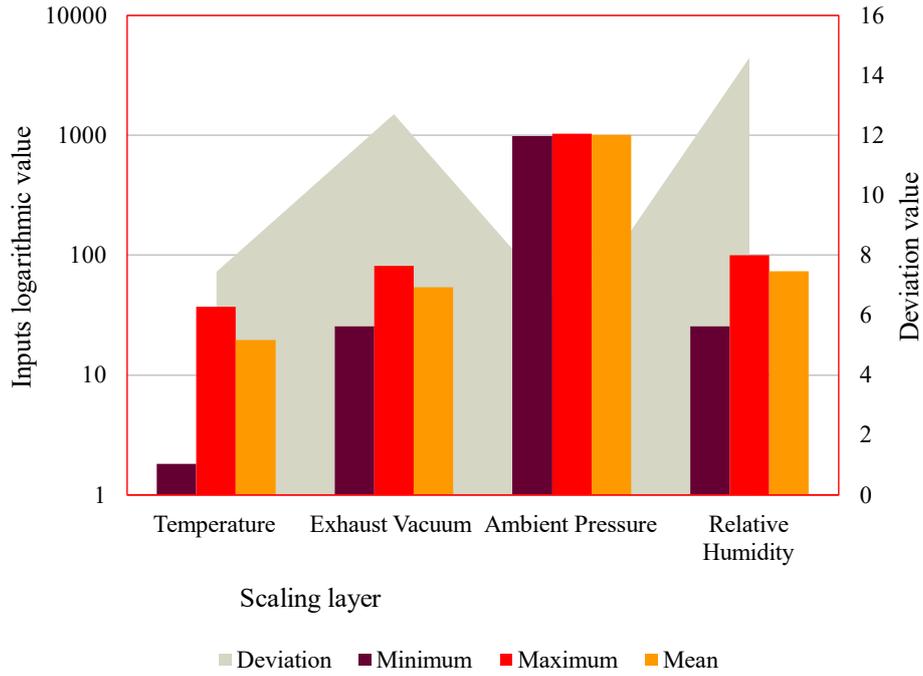

Fig. 11  The scaling layer results for the input variables with their minimum, maximum, mean, deviation, and scaler values.

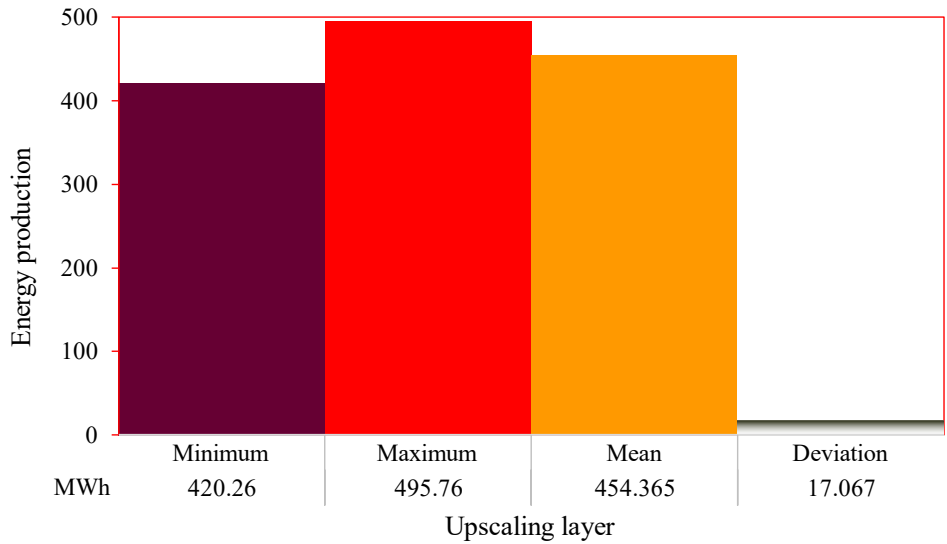

Fig. 12  Statistics of the upscaling layer and scaling factor application for the output variable "energy production" with exploring the central tendency and spread of the data.

The bounding layer is a mathematical expression that limits the output between two specific margins. It uses a formula to achieve this limitation by comparing the output value to the upper and lower bounds and adjusting it accordingly [88].

Outlier cleaning identifies and removes extreme values in a dataset significantly different from other values, usually to remove anomalies or errors that could skew results. The best parameter value for cleaning outliers depends on the specific context and goals of the analysis. Statistical tests (Z-score or modified Z-score method) and visual inspection are used to identify outliers. The Z-score is given in Eq. 18. The cleaning parameter value, ranging from 1 to 10, determines the strictness of the cleaning process, with a higher value removing more outliers but potentially valuable data and a lower value retaining more outliers but reducing accuracy. In this case study, the cleaning parameter value is considered 3, resulting in no outliers with high dataset accuracy.

$$(X - \mu) / \sigma \tag{18}$$

where $X$ is a data point, $\mu$ is the mean, and $\sigma$ is the standard deviation, which measures the number of standard deviations a value is from the mean.

The NNs parameters are established through a process of random initialization utilizing a uniform distribution with a total quantity of parameters randomly of 19 using Neural Designer. These parameters embody the starting configuration of the NNs and can later be refined through training and validation methods. The employment of random initialization is a widely accepted practice in NNs as it helps to hinder overfitting and guarantees that the network can generalize effectively to novel and unseen data.

### 5.2 Model training and appraisement

The training strategy is a vital component in the learning process, modifying and adjusting the network's parameters to attain optimal results with minimal loss, improving the network's accuracy and making more accurate predictions over time [89]. Optimization algorithms are used in training strategy to learn and adjust the parameters continuously and iteratively, to minimize the loss function, which measures the difference between predicted and actual outputs [90].

The loss index with two components of the error term and regularization term serves two primary purposes in the neural network learning process: to define the functions for the network performance and to assess the quality of the representation required for the learning process [89]. In this study, the error term (L2 regularization method) measures how accurately the network fits the data set, and the regularization term (Normalized Squared Error (MSE)) smooths the neural network's responses to prevent overfitting by evaluating parameter values and adding them to the error term.

This study used the inverse Hessian approximation method (Broyden Fletcher Goldfarb Shanno algorithm) to obtain a fitting training rate with a tolerance of 0.001 [91]. The learning rate method (Brent Method) defines the training direction. The minimum loss decrease parameter sets the minimum improvement in the loss function between two successive epochs to 0. The loss goal sets the desired value for the loss function 0.001. The maximum selection error increases parameter sets the norm of the objective function gradient's goal value to 100. The maximum number of epochs at which the selection error increases is 1000. Lastly, the maximum time parameter sets the maximum time for the training process to 1 hour.

The training and selection errors throughout each iteration are depicted in Fig. 13, which shows desirable trends of the training error initially starting at 1.04319 and decreasing to 0.06303 after 98 epochs, while the selection error initially starts at 0.58971 and falls to 0.05852 after 98 epochs. The optimization results indicate that the training process was stopped after 98 epochs due to the stopping criterion of minimum loss decreases.

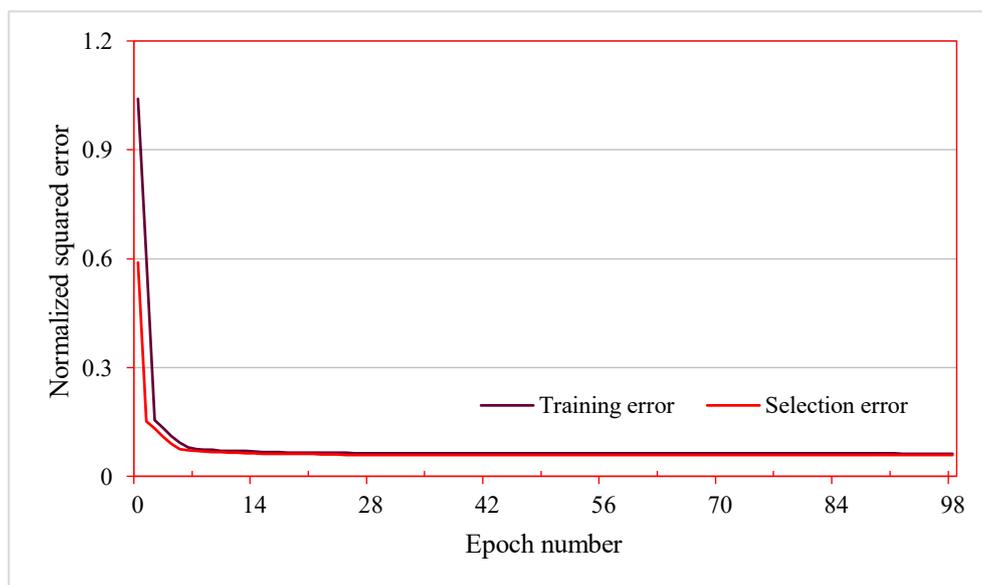

Fig. 13   Training and selection errors throughout iterations.

The growing inputs algorithm selects this application's optimal set of inputs, progressively adding inputs based on their correlations with the targets. The algorithm involves 3 trials, with a maximum of 4 inputs in the neural network and a selection error goal of 0. The algorithm stops if the selection error increases in 100 consecutive iterations or after 1000 iterations, with a maximum time of 3600 seconds. The minimum and maximum correlations considered are 0 and 1, respectively. The elapsed time for the training process is recorded as 0 seconds. The final training error and selection error values are 0.0630 and 0.0585, respectively. The growing neurons algorithm was used to select the optimal neurons for the neural network. The optimal order of the neurons was determined to be 2, with an optimum training error of 0.063301 and an optimum selection error of 0.058615. The neural network achieved these results after 10 epochs (Fig. 14), and the stopping criterion was the maximum number of neurons. The algorithm took 7 seconds to complete.

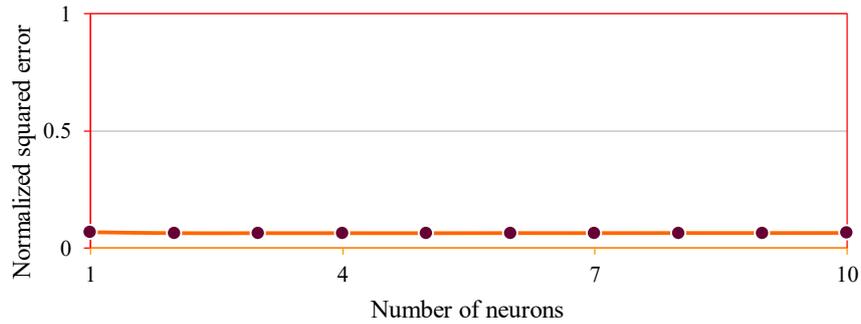

Fig. 14  The neurons' training and selection errors with a stopping criterion of the maximum number of neurons and a total runtime of 7 seconds.

The results of the growing inputs algorithm for selecting the optimal set of inputs are 4, with an optimum training error of 0.06347 and an optimum selection error of 0.059137 after 4 epochs in 2 seconds.

The analysis of training followed by the testing process aims to evaluate the goodness-of-fit of the statistical model with the set of observations using the R2 coefficient, quantifying the proportion of variation in the predicted variable from the actual values, where a perfect fit results in a value of 1. This study reports a goodness-of-fit parameter for energy production with a determination value of 0.944603. The goodness-of-fit in Fig. 15 for the target output variable displays the predicted values against the actual values. The best prediction result line indicates the outputs equaling the target variable, excluding some scaled outputs that fall outside the defined target range.

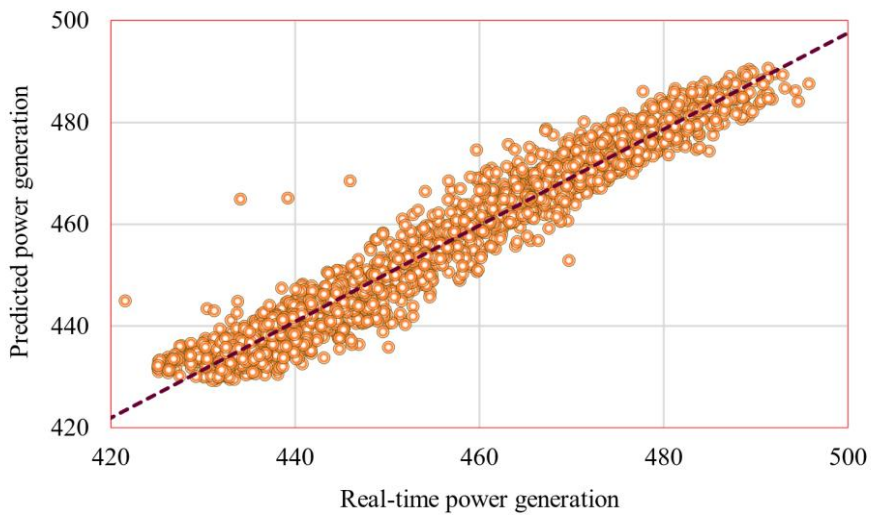

Fig. 15  Actual vs. predicted energy production (output variable) goodness-of-fit representation.

The error metrics for training, selection, and testing datasets were reported in Fig. 16, with the sum squared error being the highest for the training dataset, while the mean squared error was highest for the selected dataset. The root mean squared error was reported to be the highest for the testing dataset, and the normalized squared error was highest for the training dataset. The Minkowski error was highest for the training dataset.

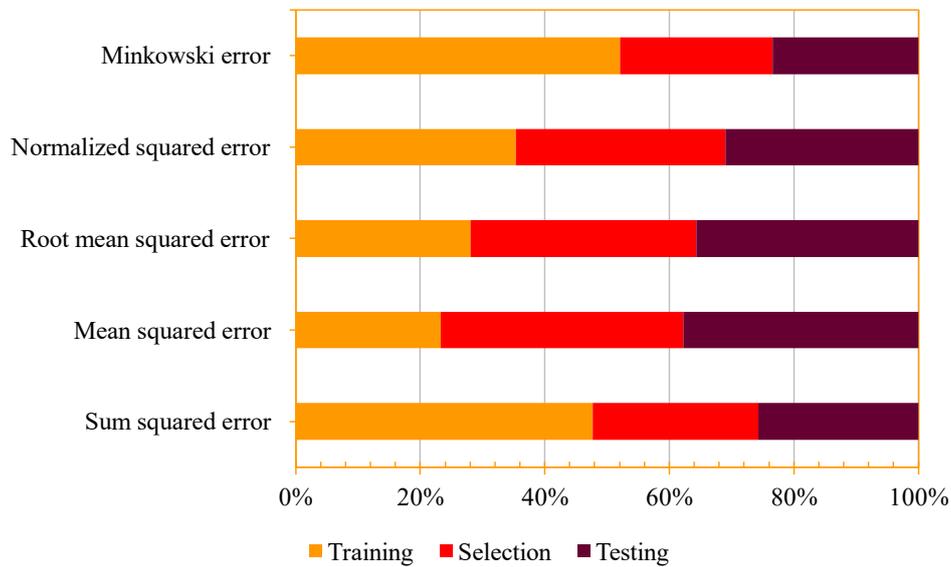

Fig. 16  The errors report for training, selection, and testing datasets analysis error metrics.

Errors statistics in Fig. 17 provide valuable insight into the quality of the model that the absolute error ranges from 0.00170898 to 30.7196, with a mean of 3.12909 and a standard deviation of 2.52692. The relative error ranges from 0.0000226 to 0.406882, with a mean of 0.0414449 and a standard deviation of 0.0334691. The percentage error ranges from 0.00226356 to 40.6882, with a mean of 4.14449 and a standard deviation of 3.34691. These insights can aid in fine-tuning and enhancing the neural network's performance assuring target loss index.

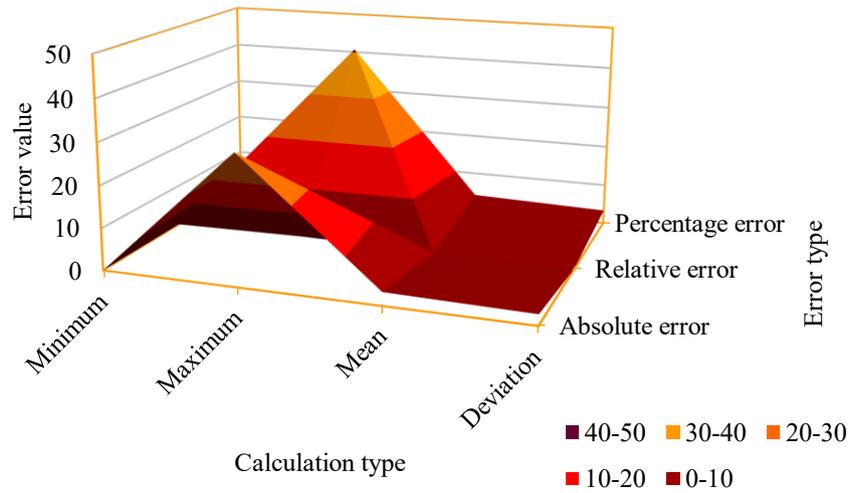

Fig. 17 The model quality appraisal in the context of errors statistics.

The relative error distribution in the neural network's predictions in Fig. 18 shows that most errors fall within the bin centered around 0, with frequencies of 29.948, gradually decreasing as the bins move away from the center. This analysis provides insight into the accuracy of the neural network's predictions and can help identify any outliers or areas for improvement.

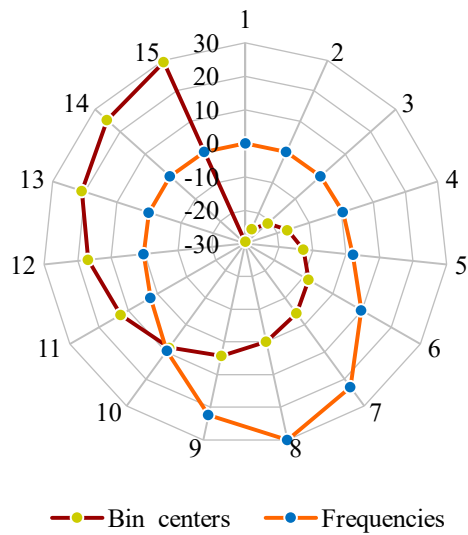

Fig. 18 Errors distribution and frequency analysis of the output variable.

The maximal errors of the output variable are reported in Fig. 19 to evaluate the difference between the predicted value and the actual value of the energy production for each sample that ranges from 11.08 to 30.72, with the highest error occurring in sample number 1. Identifying which testing samples have the largest errors permit recognizing deficiencies and potential areas for improvement in the optimization model.

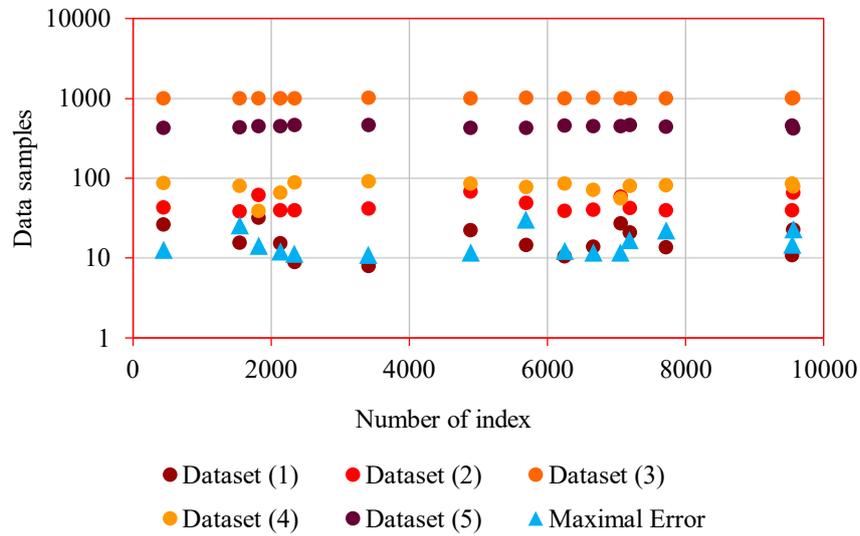

Fig. 19  Assessing the output variable maximal errors within various datasets of sample data.

The input and output variables' importance based on their influence, interactions, and values indicates positive and negative values that represent an increase or opposite in output variables, respectively. Values close to zero mean little to no change in the output variable. Despite the differences in the values in Fig. 20, both Energy and Model output-input importance agree that Exhaust Vacuum is a more important input variable than the Relative humidity and Ambient Pressure. At the same time, Temperature remained the most critical input variable for output control and improvement.

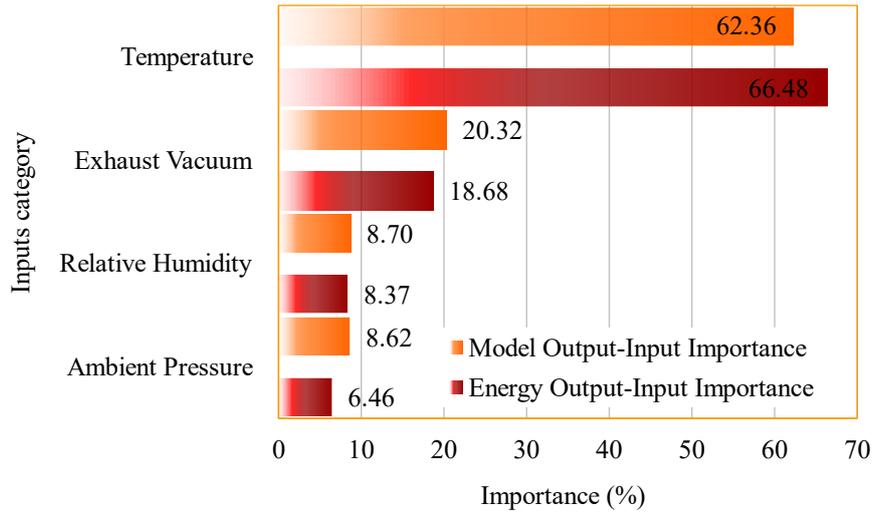

Fig. 20 Input-output variables' importance in the context of model performance and output target objective.

## 5.3 Numerical model

The numerical predictive model in Python 3.1.x (Appendix 1) can be represented as a function that utilizes the input variables to generate the output target variable, integrating it into other simulation and analysis software and tools for the production mode. The optimization employed in the model propagates information feedforward through scaling, perceptron, and unscaling layers to ensure precise predictions for the output variable that can be adopted or extended for other than the specific objective.

## 5.4 Presumptions

This study uses indices scutter plots ranging from 1 to 4 to identify the most significant variables affecting the CCPP's performance. Additionally, indices of the shown column, which range from 1 to 6, are considered to evaluate the model's performance. A minimum correlation of 0.25 is set for the variances to ensure the reliability of the results. Furthermore, the maximum error number of samples allowed is 15 to prevent the model from overfitting. The CCPP models proposed in Figs. 1-2 is developed and adapted based on thoroughly analyzing the literature concepts [92–100] relevant to the case study's requirements.

# 6 Results and Discussion

A predictive model optimizes system performance through approximation techniques using emerging statistics and machine learning methods. The resulting optimized model can forecast or estimate energy production for CCPP simply by employing input variables. However, as the target output is dependent on individuals and influenced by environmental input interrelated behaviors, obtaining optimal operating points through traditional methods is challenging without using machine learning "black-box" tools and techniques. This scenario can be expanded to evaluate the system life-cycle efficiencies (technical, technological, and ecological) [14] using a similar approach, developing a decision-making strategy that balances the influence of a plant generation on social, economic, technical, institutional, and environmental aspects on local and national levels [16]. Therefore, the numerical model is proposed to evaluate optimal operating conditions, allowing a predictive and approximation model to simulate different operating scenarios and adjust input variables to improve system output efficiency. Since environmental input variables can be easily adjusted and controlled, the proposed numerical model (Appendix 1) contributes to proper power management and generation decision-making by adjusting and combining optimal input variance values. For instance, under normal operating conditions and reported environmental parameters, the CCPP generated 452 MW. However, our findings suggest that changing some input parameters can lead to a 2.23% (462.1 MW) efficiency that ensures an output greater than or equal to 450 MW (Fig. 21). The proposed model offers the ability to simulate various operating scenarios and adjust input variables to enhance efficiency.

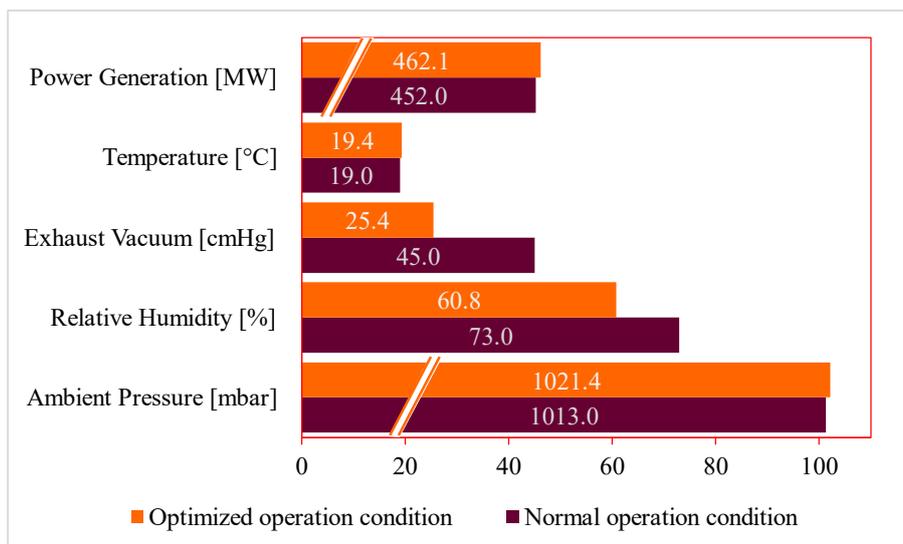

Fig. 21   Representation of CCPP performance and generation efficiency optimization.

The interdependency analysis of the input variables and target objective (output variable) is depicted in Fig. 7. The maximal correlation values among the top 4 input variables are shown in Fig. 7 (a). The variable with the highest correlation value with the target objective is Temperature (0.851175), followed by Exhaust Vacuum (-0.413502), Ambient Pressure (-0.507549), and Relative Humidity (-0.542535). At the same time, part (b) shows the correlation matrix of all input variables with each other, including the correlation values between each input variable and the target objective. The correlation coefficient values range from -1 to 1, where -1 represents a strong negative correlation, 0 represents no correlation, and 1 represents a strong positive correlation. The diagonal values of the matrix represent the correlation of each input variable with itself, which is always 1. And part (c) shows the Spearman correlation values of the top 4 input variables. The Spearman correlation is a non-parametric measure of the correlation between two variables, which evaluates how well the relationship between two variables can be described using a monotonic function. The variable with the highest Spearman correlation value with the target objective is Temperature (0.85062), followed by Exhaust Vacuum (-0.426129), Ambient Pressure (-0.51879), and Relative Humidity (-0.542706).

The Spearman correlation matrix of all input variables with each other, including the Spearman correlation values between each input variable and the target objective, are presented in Fig. 7 (d). The Spearman correlation values also range from -1 to 1, with 1 indicating a strong positive monotonic correlation, -1 indicating a strong negative monotonic correlation, and 0 indicating no monotonic correlation. In (b) and (d), the variables are arranged in a matrix format, where the variables are listed in rows and columns. The diagonal values in the matrix represent the Spearman correlation of each input variable with itself, which is always 1. The table shows that Temperature has the highest correlation with Exhaust Vacuum, followed by Relative Humidity and Ambient Pressure.

# 7 Conclusion

This study emphasizes the importance of understanding the interaction between components in combined cycle power plants (CCPP) and the interdependence of optimization objectives to increase energy efficiency, reduce emissions and improve cost-effectiveness. The proposed numerical model improves efficiency by simulating various operating scenarios and adjusting optimal parameters, significantly increasing supply efficiency. Accuracy is demonstrated through a goodness-of-fit parameter, highlighting the importance of a clean and organized dataset. The study provides a systematic tutorial on applying machine learning methods in energy systems, focusing on prediction modeling that benefits interdisciplinary researchers and readers as follows:

- The proposed numerical model enhances efficiency by simulating various operating scenarios and adjusting optimal parameters, leading to a 2.23% increase in supply efficiency (from 452 MW to 462.1 MW) by controlling environmental factors.
- The goodness-of-fit parameter for energy production with a determination value of 0.944603 shows the overall accuracy competency.
- A clean and organized data set is necessary to model data-driven systems accurately. A model should be demonstrated in the context of the intended problem with instances of input and output to evaluate performance and accuracy, which is systematically deployed in this study.
- The importance of considering a systematic approach from data analysis to interaction between different components of CCPPs and their optimization objectives to increase energy efficiency, reduce emissions, and improve cost-effectiveness through AI-coherent optimization tools and techniques is addressed.
- In addition, this study provides a systematic tutorial on applying machine learning methods in energy systems, simplifying the application process, and expanding the field for new optimization challenges. The focus on prediction modeling benefits researchers and readers from diverse interdisciplinary domains.

## Terminology

A concise terminology list is retrieved from [81,89,101] in terms of the frequency of use in this study.

1. Data science: It encompasses emerging processes, scientific methodologies, algorithms, and systems to explore patterns and insights from raw data using various mathematical, statistical, computer science, and information science tools and techniques to analyze and understand the data.

2. Machine learning: It is a type of artificial intelligence that allows machines to learn from data and improve their performance over time.

3. Algorithm: A set of instructions that enable a computer to follow and solve a problem or complete a task.

4. Artificial Neural Networks (ANNs): A computer system modeled after the structure and function of the human brain, which is capable of learning and making decisions based on complex data inputs.

5. Instance: Instance: The subject or object of which predictions are to be made.

6. Feature: A characteristic or attribute of an instance used in a prediction task.

7. Objective: A metric that the algorithm aims to optimize.

8. Label: The result of a prediction task, either the outcome produced by a machine learning system or the correct answer supplied in the training data.

9. Model: A statistical representation of a prediction task that can be trained on examples and then used to make predictions.

10. Training set: A set of examples used to train a machine learning model.

11. Test set: A set of examples used to evaluate the final performance of a machine learning model after training.

12. Validation set: A set of examples used to evaluate and tune the performance of a machine learning model during training.

13. Bias-variance tradeoff: A model's ability to fit the training data well (low bias) and its ability to generalize to unseen data (low variance).

14. Overfitting: When a model performs well on the training set but poorly on unseen data due to the model being too complex and fitting to the noise in the training data.

15. Regularization: A technique to prevent overfitting by adding a penalty term to the objective function the algorithm tries to optimize.

16. Metric: A numerical value that is important but not constantly directly optimized.

17. Precision: A metric that measures the proportion of true positive predictions among all positive predictions.

18. Recall: A metric that evaluates the proportion of true positive predictions among all actual positive instances.

19. F1 Score: A metric that combines precision and recall, taking into account both the false positives and false negatives.

20. Confusion Matrix: A table that visualizes the performance of a classification model, showing the number of true positives, false positives, true negatives, and false negatives.

21. ROC Curve: A graphical representation of the performance of a binary classification model, plotting the true positive rate against the false positive rate at different threshold settings.

22. AUC: A metric that measures the area under the ROC curve, indicating the overall performance of a binary classification model.

23. Cross-Validation: A technique for assessing the performance of a machine learning model by dividing the training data into several subsets, training the model on different subsets, and evaluating it on the remaining subsets.

24. Hyperparameter: Parameters that are set before training, as opposed to parameters learned during training, such as learning rate, number of layers in NNs, regularization strength, number of decision trees in a random forest, searching for the best combination of parameters to improve the model's performance, leading to better performance, faster training and improved generalization to unseen data.

25. Feature column: A group of related features in a data table (dataset).

26. Example: An instance (with its associated features) and a label.

27. Pipeline: The infrastructure surrounding a machine learning algorithm, including data collection, preparation of training data, model training, and deployment to production.

28. Indices scatter plot: A range of indices (or columns) in a dataset that are being plotted against each other in a scatter plot.

29. Indices of shown column: A range of indices (or columns) in a dataset that are being displayed or analyzed.

30. Minimum correlation of the variances: A threshold value for the correlation between the variances of different variables in a dataset to ensure that the variances of different variables are not strongly correlated with each other in order to avoid issues with multicollinearity.

31. Maximum error number of sample: A limit on the number of errors that are allowed in an instance.

# Appendix

Appendix 1: The proposed numerical calculation of the CCPP optimum generation capacity in python.

The provided code assumes that the input variables consist of a single set of four optimum input variables. By using multiple input datasets, the system performance and output efficiency can be evaluated accordingly. Advanced methods such as loops or vectorization can be employed for this purpose.

```python
#The proposed numerical calculation of the CCPP optimum generation
capacity in python. Using a neural network model, this Python code
calculates the optimal Combined Cycle Power Plant (CCPP)
generation capacity. The input data, including temperature,
evaporation, atmospheric pressure, and relative humidity, are
initially scaled using predefined scaling factors and offsets. The
neural network comprises two perceptron layers with weights and
biases, which compute the outputs of the first layer and the final
production. The output is subsequently unscaled using an unscaling
factor and offset. The optimal CCPP power generation is calculated
and displayed based on the provided input values, resulting in a
power generation of at least 462 MW.

import numpy as np

# Scaling values
temp_scale = 7.452469826
temp_offset = 19.65119934
ev_scale = 12.70790005
ev_offset = 54.30580139
ap_scale = 5.938789845
ap_offset = 1013.26001
rh_scale = 14.60029984
rh_offset = 73.30899811

# Perceptron weights and biases
p1_w0 = np.array([0.593324, 0.00657032, 0.0933036, 0.0310159])
p1_b0 = 0.688402
p1_w1 = np.array([0.288555, 0.204765, -0.144645, 0.070106])
p1_b1 = -0.110117
p2_w0 = np.array([-1.34001, -1.13091])
p2_b0 = 0.582504

# Unscaling factors
us_scale = 17.06699944
us_offset = 454.3649902

# Function to scale the input data
def scale_input_data(temp, ev, ap, rh):
    scaled_temp = (temp - temp_offset) / temp_scale
    scaled_ev = (ev - ev_offset) / ev_scale
    scaled_ap = (ap - ap_offset) / ap_scale
    scaled_rh = (rh - rh_offset) / rh_scale
    return scaled_temp, scaled_ev, scaled_ap, scaled_rh

# Function to unscale the output data
def unscale_output_data(output):
    unscaled_output = output * us_scale + us_offset
    return unscaled_output

# Function to calculate the output
def neural_network_output(temp, ev, ap, rh):
    # Scale the input data
```

```python
    scaled_temp, scaled_ev, scaled_ap, scaled_rh = 
scale_input_data(temp, ev, ap, rh)

    # Outputs of the first perceptron layer
    p1_output0 = np.tanh(p1_b0 + np.dot(np.array([scaled_temp, 
scaled_ev, scaled_ap, scaled_rh]), p1_w0))
    p1_output1 = np.tanh(p1_b1 + np.dot(np.array([scaled_temp, 
scaled_ev, scaled_ap, scaled_rh]), p1_w1))

    # Output of the second perceptron layer
    p2_output0 = p2_b0 + np.dot(np.array([p1_output0, 
p1_output1]), p2_w0)

    # Unscale the output data
    unscaled_output = unscale_output_data(p2_output0)

    return unscaled_output

# Optimal values (Power Generation        ≥ 462 [MW])
temp = 19.4
ev = 25.4
ap = 1021.4
rh = 60.8

plant_output = neural_network_output(temp, ev, ap, rh)
print(f"The CCPP power generation at optimal operating condition 
[MW]: {plant_output:.2f}")
```

# 8 Acknowledgments

We would like to sincerely express our gratitude to all who (anonymously) contributed to completing this article. Your unwavering support, insightful comments, and invaluable guidance, especially during critical moments, have played an instrumental role in shaping this work. We are deeply grateful for your contributions and appreciate your willingness to join this journey. Also, we acknowledge the use of various simulation, word processing, contents management, and visualization web-based and local interfaces software, e.g., Python, Zotero, Source Code Beautifier, Neural Designer Illustrator, etc.

# 9 References

[1]    Pattanayak L, Padhi BN. Thermodynamic simulation and economic analysis of combined cycle with inlet air cooling and fuel pre-heating: Performance enhancement and emission reduction. Energy Convers Manag 2022;267:115884. https://doi.org/10.1016/j.enconman.2022.115884.

[2]   Skalska K, Miller JS, Ledakowicz S. Trends in NO$_x$ abatement: A review. Sci Total Environ 2010;408:3976–89. https://doi.org/10.1016/j.scitotenv.2010.06.001.
[3]   Iliyas SA, Elshafei Moustafa, Habib MA, Adeniran AA. RBF neural network inferential sensor for process emission monitoring. Control Eng Pract 2013;21:962–70. https://doi.org/10.1016/j.conengprac.2013.01.007.
[4]   European Environment Agency. Emissions and energy use in large combustion plants in Europe. Environ Indic 2022. https://www.eea.europa.eu/ims/emissions-and-energy-use-in (accessed August 14, 2022).
[5]   KAYA H, TÜFEKCİ P, UZUN E. Predicting CO and NO$_x$ emissions from gas turbines: novel data and a benchmark PEMS. Turk J Electr Eng Comput Sci 2019;27:4783–96. https://doi.org/10.3906/elk-1807-87.
[6]   Huang G-B, Zhu Q-Y, Siew C-K. Extreme learning machine: Theory and applications. Neurocomputing 2006;70:489–501. https://doi.org/10.1016/j.neucom.2005.12.126.
[7]   Huang G-B, Zhou H, Ding X, Zhang R. Extreme Learning Machine for Regression and Multiclass Classification. IEEE Trans Syst Man Cybern Part B Cybern 2012;42:513–29. https://doi.org/10.1109/TSMCB.2011.2168604.
[8]   Saleel CA. Forecasting the energy output from a combined cycle thermal power plant using deep learning models. Case Stud Therm Eng 2021;28:101693. https://doi.org/10.1016/j.csite.2021.101693.
[9]   Talaat M, Gobran MH, Wasfi M. A hybrid model of an artificial neural network with thermodynamic model for system diagnosis of electrical power plant gas turbine. Eng Appl Artif Intell 2018;68:222–35. https://doi.org/10.1016/j.engappai.2017.10.014.
[10]  Ibrahim TK, Mohammed MK, Awad OI, Abdalla AN, Basrawi F, Mohammed MN, et al. A comprehensive review on the exergy analysis of combined cycle power plants. Renew Sustain Energy Rev 2018;90:835–50. https://doi.org/10.1016/j.rser.2018.03.072.
[11]  Sun S, Fu J, Zhu F, Du D. A hybrid structure of an extreme learning machine combined with feature selection, signal decomposition and parameter optimization for short-term wind speed forecasting. Trans Inst Meas Control 2020;42:3–21. https://doi.org/10.1177/0142331218771141.
[12]  Aldeneh Z, Mower Provost E. You're Not You When You're Angry: Robust Emotion Features Emerge by Recognizing Speakers. IEEE Trans Affect Comput 2021:1–1. https://doi.org/10.1109/TAFFC.2021.3086050.
[13]  Yadav SP, Zaidi S, Mishra A, Yadav V. Survey on Machine Learning in Speech Emotion Recognition and Vision Systems Using a Recurrent Neural Network (RNN). Arch Comput Methods Eng 2022;29:1753–70. https://doi.org/10.1007/s11831-021-09647-x.
[14]  Danish MSS, Senjyu T, Zaheb H, Sabory NR, Ibrahimi AM, Matayoshi H. A novel transdisciplinary paradigm for municipal solid waste to energy. J Clean Prod 2019;233:880–92. https://doi.org/10.1016/j.jclepro.2019.05.402.
[15]  Waas T, Hugé J, Block T, Wright T, Benitez-Capistros F, Verbruggen A. Sustainability Assessment and Indicators: Tools in a Decision-Making Strategy for Sustainable Development. Sustainability 2014;6:5512–34. https://doi.org/10.3390/su6095512.
[16]  Danish MSS, Senjyu T, Ibrahimi AM, Ahmadi M, Howlader AM. A managed framework for energy-efficient building. J Build Eng 2019;21:120–8. https://doi.org/10.1016/j.jobe.2018.10.013.
[17]  Danish MSS, Sabory NR, Ibrahimi AM, Senjyu T, Ahadi MH, Stanikzai MZ. A Concise Overview of Energy Development Within Sustainability Requirements. In: Danish MSS, Senjyu T, Sabory NR, editors. Sustain. Outreach Dev. Ctries., Singapore: Springer; 2021, p. 15–27. https://doi.org/10.1007/978-981-15-7179-4_2.
[18]  Danish MSS, Senjyu TS. Green Building Efficiency and Sustainability Indicators. Green Build. Manag. Smart Autom. 1st ed., Pennsylvania, United States: IGI Global; 2020, p. 128–45. https://doi.org/10.4018/978-1-5225-9754-4.ch006.
[19]  Korpela T, Kumpulainen P, Majanne Y, Häyrinen A. Model based NO$_x$ emission monitoring in natural gas fired hot water boilers. IFAC-Pap 2015;48:385–90. https://doi.org/10.1016/j.ifacol.2015.12.409.
[20]  Lin T-Y, Chiu Y-H, Lin Y-N, Chang T-H, Lin P-Y. Greenhouse gas emission indicators, energy consumption efficiency, and optimal carbon emission allowance allocation of the EU countries in 2030. Gas Sci Eng 2023;110:204902. https://doi.org/10.1016/j.jgsce.2023.204902.
[21]  Danish MSS, Senjyu T, Funabashia T, Ahmadi M, Ibrahimi AM, Ohta R, et al. A sustainable microgrid: A sustainability and management-oriented approach. Energy Procedia 2019;159:160–7. https://doi.org/10.1016/j.egypro.2018.12.045.
[22]  Tarmanini C, Sarma N, Gezegin C, Ozgonenel O. Short term load forecasting based on ARIMA and ANN approaches. Energy Rep 2023;9:550–7. https://doi.org/10.1016/j.egyr.2023.01.060.
[23]  Bansal M, Goyal A, Choudhary A. A comparative analysis of K-Nearest Neighbor, Genetic, Support Vector Machine, Decision Tree, and Long Short Term Memory algorithms in machine learning. Decis Anal J 2022;3:100071. https://doi.org/10.1016/j.dajour.2022.100071.
[24]  Mhlanga D. Artificial Intelligence and Machine Learning for Energy Consumption and Production in Emerging Markets: A Review. Energies 2023;16:745. https://doi.org/10.3390/en16020745.
[25]  Xu Y, Klein B, Li G, Gopaluni B. Evaluation of logistic regression and support vector machine approaches for XRF based particle sorting for a copper ore. Miner Eng 2023;192:108003. https://doi.org/10.1016/j.mineng.2023.108003.

[26] Sarker IH. Deep Learning: A Comprehensive Overview on Techniques, Taxonomy, Applications and Research Directions. SN Comput Sci 2021;2:420. https://doi.org/10.1007/s42979-021-00815-1.
[27] Entezari A, Aslani A, Zahedi R, Noorollahi Y. Artificial intelligence and machine learning in energy systems: A bibliographic perspective. Energy Strategy Rev 2023;45:101017. https://doi.org/10.1016/j.esr.2022.101017.
[28] Pai P-F, Chen T-C. Rough set theory with discriminant analysis in analyzing electricity loads. Expert Syst Appl 2009;36:8799–806. https://doi.org/10.1016/j.eswa.2008.11.012.
[29] Busari GA, Lim DH. Crude oil price prediction: A comparison between AdaBoost-LSTM and AdaBoost-GRU for improving forecasting performance. Comput Chem Eng 2021;155:107513. https://doi.org/10.1016/j.compchemeng.2021.107513.
[30] Shi Z, Zhu J, Wei H. SARSA-based delay-aware route selection for SDN-enabled wireless-PLC power distribution IoT. Alex Eng J 2022;61:5795–803. https://doi.org/10.1016/j.aej.2021.11.029.
[31] Steinwandter V, Borchert D, Herwig C. Data science tools and applications on the way to Pharma 4.0. Drug Discov Today 2019;24:1795–805. https://doi.org/10.1016/j.drudis.2019.06.005.
[32] Tebenkov E, Prokhorov I. Machine learning algorithms for teaching AI chat bots. Procedia Comput Sci 2021;190:735–44. https://doi.org/10.1016/j.procs.2021.06.086.
[33] Daradkeh M, Abualigah L, Atalla S, Mansoor W. Scientometric Analysis and Classification of Research Using Convolutional Neural Networks: A Case Study in Data Science and Analytics. Electronics 2022;11:2066. https://doi.org/10.3390/electronics11132066.
[34] Chawla Y, Shimpo F, Sokołowski MM. Artificial intelligence and information management in the energy transition of India: lessons from the global IT heart. Digit Policy Regul Gov 2022;24:17–29. https://doi.org/10.1108/DPRG-05-2021-0062.
[35] Danish MSS, Elsayed MEL, Ahmadi M, Senjyu T, Karimy H, Zaheb H. A strategic-integrated approach for sustainable energy deployment. Energy Rep 2020;6:40–4. https://doi.org/10.1016/j.egyr.2019.11.039.
[36] Sala S, Ciuffo B, Nijkamp P. A systemic framework for sustainability assessment. Ecol Econ 2015;119:314–25. https://doi.org/10.1016/j.ecolecon.2015.09.015.
[37] Tolentino-Zondervan F, Bogers E, van de Sande L. A Managerial and Behavioral Approach in Aligning Stakeholder Goals in Sustainable Last Mile Logistics: A Case Study in the Netherlands. Sustainability 2021;13:4434. https://doi.org/10.3390/su13084434.
[38] Wei C, Zhang C, Vardar O, Watson J, Canbulat I. Quantitative assessment of energy changes in underground coal excavations using numerical approach. Geohazard Mech 2022. https://doi.org/10.1016/j.ghm.2022.11.003.
[39] Raza MA, Khatri KL, Ul Haque MI, Shahid M, Rafique K, Waseer TA. Holistic and scientific approach to the development of sustainable energy policy framework for energy security in Pakistan. Energy Rep 2022;8:4282–302. https://doi.org/10.1016/j.egyr.2022.03.044.
[40] Zhang X, Wu Y, Zhang W, Wang Q, Wang A. System Performance and Pollutant Emissions of Micro Gas Turbine Combined Cycle in Variable Fuel Type Cases. Energies 2022;15:9113. https://doi.org/10.3390/en15239113.
[41] Tomlinson LO, McCullough S. Single-Shaft Combined-Cycle Power Generation System. GE Power Syst 1998;GER-3767C:1–22.
[42] Boyce MP. 1 - Combined cycle power plants. In: Rao AD, editor. Comb. Cycle Syst. -Zero Emiss. Power Gener., Woodhead Publishing; 2012, p. 1–43. https://doi.org/10.1533/9780857096180.1.
[43] Feldmuller A, Zimmerer T, Roehr F. From Base to Cycling Operation: Innovative Operational Concepts for CCPPs. Erlangen, Germany: Siemens Energy Global GmbH & Co. KG; 2015.
[44] Wang S, Zhang G, Fu Z. Performance analysis of a novel integrated solar combined cycle with inlet air heating system. Appl Therm Eng 2019;161:114010. https://doi.org/10.1016/j.applthermaleng.2019.114010.
[45] Pashchenko D. Performance evaluation of a combined power generation system integrated with thermochemical exhaust heat recuperation based on steam methane reforming. Int J Hydrog Energy 2023;48:5823–35. https://doi.org/10.1016/j.ijhydene.2022.11.186.
[46] Razak AMY. 3 - Complex gas turbine cycle. In: Razak AMY, editor. Ind. Gas Turbines, Woodhead Publishing; 2007, p. 60–97. https://doi.org/10.1533/9781845693404.1.60.
[47] Winterbone DE, Turan A. Chapter 17 - Gas Turbines. In: Winterbone DE, Turan A, editors. Adv. Thermodyn. Eng. Second Ed., Boston: Butterworth-Heinemann; 2015, p. 381–422. https://doi.org/10.1016/B978-0-444-63373-6.00017-4.
[48] Dev N, Samsher, Kachhwaha SS, Attri R. GTA modeling of combined cycle power plant efficiency analysis. Ain Shams Eng J 2015;6:217–37. https://doi.org/10.1016/j.asej.2014.08.002.
[49] Pattanayak L, Padhi BN. Thermodynamic analysis of combined cycle power plant using regasification cold energy from LNG terminal. Energy 2018;164:1–9. https://doi.org/10.1016/j.energy.2018.08.187.
[50] Mohamed O, Khalil A. Progress in Modeling and Control of Gas Turbine Power Generation Systems: A Survey. Energies 2020;13:2358. https://doi.org/10.3390/en13092358.
[51] Taha S, IsmaiL F, ThiRuchelvam S. Gas Turbine Performance Monitoring and Operation Challenges: A Review. GAZI Univ J Sci 2022;36:154–71. https://doi.org/10.35378/gujs.948875.

[52]	Birt J. 7 Types of Observational Studies (With Examples). Indeed Career Guide 2022. https://www.indeed.com/career-advice/career-development/types-of-observational-studies (accessed September 22, 2022).
[53]	Matharu Y. An Introduction to Neural Networks. Medium 2021. https://becominghuman.ai/an-introduction-to-neural-networks-50adc4029b09 (accessed April 28, 2022).
[54]	Ehmer M, Khan F. A Comparative Study of White Box, Black Box and Grey Box Testing Techniques. Int J Adv Comput Sci Appl 2012;3:1–15. https://doi.org/10.14569/IJACSA.2012.030603.
[55]	Brahma S, Kavasseri R, Cao H, Chaudhuri NR, Alexopoulos T, Cui Y. Real-Time Identification of Dynamic Events in Power Systems Using PMU Data, and Potential Applications—Models, Promises, and Challenges. IEEE Trans Power Deliv 2017;32:294–301. https://doi.org/10.1109/TPWRD.2016.2590961.
[56]	Ardakanian O, Wong VWS, Dobbe R, Low SH, von Meier A, Tomlin CJ, et al. On Identification of Distribution Grids. IEEE Trans Control Netw Syst 2019;6:950–60. https://doi.org/10.1109/TCNS.2019.2891002.
[57]	Bellizio F, Karagiannopoulos S, Aristidou P, Hug G. Optimized Local Control for Active Distribution Grids using Machine Learning Techniques. 2018 IEEE Power Energy Soc. Gen. Meet. PESGM, 2018, p. 1–5. https://doi.org/10.1109/PESGM.2018.8586079.
[58]	Zienkiewicz AK, Ladu F, Barton DAW, Porfiri M, Bernardo MD. Data-driven modelling of social forces and collective behaviour in zebrafish. J Theor Biol 2018;443:39–51. https://doi.org/10.1016/j.jtbi.2018.01.011.
[59]	Dobbe R, Sondermeijer O, Fridovich-Keil D, Arnold D, Callaway D, Tomlin C. Toward Distributed Energy Services: Decentralizing Optimal Power Flow With Machine Learning. IEEE Trans Smart Grid 2020;11:1296–306. https://doi.org/10.1109/TSG.2019.2935711.
[60]	Deka D, Backhaus S, Chertkov M. Structure Learning in Power Distribution Networks. IEEE Trans Control Netw Syst 2018;5:1061–74. https://doi.org/10.1109/TCNS.2017.2673546.
[61]	Liao Y, Weng Y, Liu G, Rajagopal R. Urban MV and LV Distribution Grid Topology Estimation via Group Lasso. IEEE Trans Power Syst 2019;34:12–27. https://doi.org/10.1109/TPWRS.2018.2868877.
[62]	Sun M, Cremer J, Strbac G. A novel data-driven scenario generation framework for transmission expansion planning with high renewable energy penetration. Appl Energy 2018;228:546–55. https://doi.org/10.1016/j.apenergy.2018.06.095.
[63]	Zhao C, Guan Y. Data-Driven Stochastic Unit Commitment for Integrating Wind Generation. IEEE Trans Power Syst 2016;31:2587–96. https://doi.org/10.1109/TPWRS.2015.2477311.
[64]	Duan C, Jiang L, Fang W, Liu J. Data-Driven Affinely Adjustable Distributionally Robust Unit Commitment. IEEE Trans Power Syst 2018;33:1385–98. https://doi.org/10.1109/TPWRS.2017.2741506.
[65]	Jokar P, Arianpoo N, Leung VCM. Electricity Theft Detection in AMI Using Customers' Consumption Patterns. IEEE Trans Smart Grid 2016;7:216–26. https://doi.org/10.1109/TSG.2015.2425222.
[66]	Glavic M, Fonteneau R, Ernst D. Reinforcement Learning for Electric Power System Decision and Control: Past Considerations and Perspectives. IFAC-Pap 2017;50:6918–27. https://doi.org/10.1016/j.ifacol.2017.08.1217.
[67]	Ernst D, Glavic M, Wehenkel L. Power systems stability control: reinforcement learning framework. IEEE Trans Power Syst 2004;19:427–35. https://doi.org/10.1109/TPWRS.2003.821457.
[68]	Karagiannopoulos S, Aristidou P, Hug G. Data-Driven Local Control Design for Active Distribution Grids Using Off-Line Optimal Power Flow and Machine Learning Techniques. IEEE Trans Smart Grid 2019;10:6461–71. https://doi.org/10.1109/TSG.2019.2905348.
[69]	Xiong R, Cao J, Yu Q. Reinforcement learning-based real-time power management for hybrid energy storage system in the plug-in hybrid electric vehicle. Appl Energy 2018;211:538–48. https://doi.org/10.1016/j.apenergy.2017.11.072.
[70]	Lesage-Landry A, Taylor JA. Setpoint Tracking With Partially Observed Loads. IEEE Trans Power Syst 2018;33:5615–27. https://doi.org/10.1109/TPWRS.2018.2804353.
[71]	van der Linden I, Haned H, Kanoulas E. Global Aggregations of Local Explanations for Black Box models 2019.
[72]	Goodfellow I, Bengio Y, Courville A. Deep Learning. Cambridge, Massachusetts: The MIT Press; 2016.
[73]	Cheng L, Wang Z, Jiang F, Zhou C. Real-Time Optimal Control for Spacecraft Orbit Transfer via Multiscale Deep Neural Networks. IEEE Trans Aerosp Electron Syst 2019;55:2436–50. https://doi.org/10.1109/TAES.2018.2889571.
[74]	Silvestrini S, Lavagna M. Neural-aided GNC reconfiguration algorithm for distributed space system: development and PIL test. Adv Space Res 2021;67:1490–505. https://doi.org/10.1016/j.asr.2020.12.014.
[75]	Montague PR. Reinforcement Learning: An Introduction, by Sutton, R.S. and Barto, A.G. Trends Cogn Sci 1999;3:360. https://doi.org/10.1016/S1364-6613(99)01331-5.
[76]	Goodfellow I, Pouget-Abadie J, Mirza M, Xu B, Warde-Farley D, Ozair S, et al. Generative adversarial networks. Commun ACM 2020;63:139–44. https://doi.org/10.1145/3422622.
[77]	Silvestrini S, Lavagna M. Deep Learning and Artificial Neural Networks for Spacecraft Dynamics, Navigation and Control. Drones 2022;6:270. https://doi.org/10.3390/drones6100270.
[78]	Yu W. A Quasi-Newton Method for Estimating the Parameter in a Nonlinear Hyperbolic System. J Math Anal Appl 1999;231:397–424. https://doi.org/10.1006/jmaa.1998.6227.

[79] Goswami A. Combined Cycle Power Plant Data Set, University of California, Irvine 2020.
[80] Tüfekci P. Prediction of full load electrical power output of a base load operated combined cycle power plant using machine learning methods. Int J Electr Power Energy Syst 2014;60:126–40. https://doi.org/10.1016/j.ijepes.2014.02.027.
[81] Alpaydin E. Introduction to Machine Learning. second edition. Cambridge, Mass: The MIT Press; 2009.
[82] Brownlee J. How to Use StandardScaler and MinMaxScaler Transforms in Python. Mach Learn Mastery 2020. https://machinelearningmastery.com/standardscaler-and-minmaxscaler-transforms-in-python/ (accessed November 14, 2022).
[83] Kuhn M, Johnson K. Applied Predictive Modeling. 1st ed. 2013, Corr. 2nd printing 2018 edition. New York: Springer; 2013.
[84] Jia X, Gong X, Liu X, Zhao X, Meng H, Dong Q, et al. Deep Sequence Learning for Prediction of Daily $NO_2$ Concentration in Coastal Cities of Northern China. Atmosphere 2023;14:467. https://doi.org/10.3390/atmos14030467.
[85] Gupta A. Spearman's Rank Correlation: The Definitive Guide To Understand. Simplilearn 2022. https://www.simplilearn.com/tutorials/statistics-tutorial/spearmans-rank-correlation (accessed October 18, 2022).
[86] Sharma S. Activation Functions in Neural Networks. Medium 2017. https://towardsdatascience.com/activation-functions-neural-networks-1cbd9f8d91d6 (accessed November 20, 2022).
[87] Bakr MH, Negm MH. Modeling and Design of High-Frequency Structures Using Artificial Neural Networks and Space Mapping. Adv. Imaging Electron Phys., vol. 174, Elsevier; 2012, p. 223–60. https://doi.org/10.1016/B978-0-12-394298-2.00003-X.
[88] Neural networks tutorial: Neural network | Neural Designer. Neural Des 2022. https://www.neuraldesigner.com/learning/tutorials/neural-network#BoundingLayer (accessed August 12, 2022).
[89] Goodfellow I, Bengio Y, Courville A. Deep Learning: Adaptive Computation and Machine Learning series. Illustrated edition. Cambridge, Massachusetts: The MIT Press; 2016.
[90] Schreiber T, Netsch C, Eschweiler S, Wang T, Storek T, Baranski M, et al. Application of data-driven methods for energy system modelling demonstrated on an adaptive cooling supply system. Energy 2021;230:120894. https://doi.org/10.1016/j.energy.2021.120894.
[91] Nocedal J. Updating quasi-Newton matrices with limited storage. Math Comput 1980;35:773–82. https://doi.org/10.1090/S0025-5718-1980-0572855-7.
[92] Vakkilainen EK. 1 - Principles of Steam Generation. In: Vakkilainen EK, editor. Steam Gener. Biomass, Butterworth-Heinemann; 2017, p. 1–17. https://doi.org/10.1016/B978-0-12-804389-9.00001-0.
[93] Ahmed ASE, Elhosseini MA, Arafat Ali H. Modelling and practical studying of heat recovery steam generator (HRSG) drum dynamics and approach point effect on control valves. Ain Shams Eng J 2018;9:3187–96. https://doi.org/10.1016/j.asej.2018.06.004.
[94] Taimoor AA, Siddiqui ME, Abdel Aziz SS. Thermodynamic Analysis of Partitioned Combined Cycle using Simple Gases. Appl Sci 2019;9:4190. https://doi.org/10.3390/app9194190.
[95] Moosazadeh Moosavi SA, Mafi M, Kaabi Nejadian A, Salehi G, Torabi Azad M. A new method to boost performance of heat recovery steam generators by integrating pinch and exergy analyses. Adv Mech Eng 2018;10:1687814018777879. https://doi.org/10.1177/1687814018777879.
[96] Almajali M, Quran OA. Parametric Study on the Performance of Combined Power Plant of Steam and Gas Turbines. J Therm Sci Eng Appl 2021;13:in progress. https://doi.org/10.1115/1.4049753].
[97] Elhosseini MA, El-Din AS, Ali HA, Abraham A. Heat recovery steam generator (HRSG) three-element drum level control utilizing Fractional order PID and fuzzy controllers. ISA Trans 2022;122:281–93. https://doi.org/10.1016/j.isatra.2021.04.035.
[98] Carazas FJG, Souza GFM de. Availability Analysis of Heat Recovery Steam Generators used in Combined Cycle Thermoelectric Power Plants. Proc. COBEM 2007, Brasília, Brazil: ABCM; 2007, p. 1–9.
[99] Zeng J, Xing M, Hou M, England GC, Yan J. How best management practices affect emissions in gas turbine power plants—An important factor to consider when strengthening emission standards. J Air Waste Manag Assoc 2018;68:945–57. https://doi.org/10.1080/10962247.2018.1460634.
[100] Jürke S. The new ATP family — The optimum turbine for every application. Fuel Energy Abstr 1996;37:452. https://doi.org/10.1016/S0140-6701(97)83816-7.
[101] Utagawa H. Machine learning glossary: The beginner's guide to machine learning. Washington DC, USA: Amazon Services LLC; 2020.